\documentclass{article}

\PassOptionsToPackage{numbers, compress}{natbib}
  \usepackage[preprint]{neurips_2026}

\usepackage[utf8]{inputenc} 
\usepackage[T1]{fontenc}    
\usepackage{hyperref}       
\usepackage{url}            
\usepackage{booktabs}       
\usepackage{amsfonts}       
\usepackage{nicefrac}       
\usepackage{microtype}      
\usepackage{xcolor}         
\usepackage{comment}
\usepackage{pifont}
\usepackage{makecell}
\newcommand{\cmark}{\ding{51}} 
\newcommand{\xmark}{\ding{55}} 
\usepackage{graphicx} 
\usepackage{amsmath}
\usepackage{algorithm}
\usepackage{algpseudocode}
\usepackage{multirow}
\usepackage{booktabs} 
\usepackage{subcaption}
\usepackage{placeins}
\usepackage{listings}
\usepackage{xcolor}
\usepackage{threeparttable}
\usepackage[table]{xcolor}
\usepackage{fontawesome5}
\usepackage{amssymb}
\usepackage{cancel}

\usepackage[most]{tcolorbox}
\tcbset{
  breakable,          
  enhanced,
  colback=gray!10,
  colframe=black!75,
  boxrule=0.5pt,
  arc=2mm
}

\lstdefinelanguage{json}{
    morestring=[b]",
    morecomment=[l]{//},
    morekeywords={true,false,null},
    sensitive=false,
}

\usepackage{tikz}
\usetikzlibrary{calc}

\newcommand{\halfcheck}[1][0.25]{%
  \tikz[baseline=-0.5ex]{
    \node[inner sep=0pt] (c) {\cmark};
    \draw[line width=0.9pt]
      ($(c.north west)!0.5-#1!(c.south east)$) --
      ($(c.north west)!0.5+#1!(c.south east)$);
  }
}

\title{ROAR: Benchmarking LMM Agents under Real-Time Constraints in Multiplayer Action-Shooter Games}

\author{%
  Dawei Wang\textsuperscript{1}\thanks{Correspondence to: Dawei Wang \texttt{<d.wang28@newcastle.ac.uk>}}
  \hspace{0.1cm} Di Zhao\textsuperscript{2}
  \hspace{0.01cm} Jianran Yu\textsuperscript{1}
  \hspace{0.01cm} Chengming Zhou\textsuperscript{3}
  \hspace{0.01cm} Marci Chi Ma\textsuperscript{1}
  \hspace{0.01cm} Xiaoyang Liu\textsuperscript{1}
  \AND Beining Zhang\textsuperscript{1}
  \hspace{0.01cm} Shuliang Chu\textsuperscript{1}
  \hspace{0.01cm} Yun Sing Koh\textsuperscript{2}
  \hspace{0.01cm} Gary Ushaw\textsuperscript{1}
  \hspace{0.01cm} Richard Davison\textsuperscript{1}
  \\
  \\
  \textsuperscript{1}Newcastle University
 \hspace{1cm} \textsuperscript{2}University of Auckland
 \hspace{1cm} \textsuperscript{3}CTW Inc.
 \\
 \\
 \href{https://github.com/tb6147877/ROAR.git}{\faGithub\ GitHub}
}

\begin{document}

\maketitle

\begin{abstract}
  Large multimodal models (LMMs) are increasingly deployed as agents that perceive, reason, and act in interactive multi-agent environments. Yet most existing benchmarks for LMM-based agents in multi-agent settings remain turn-based: the environment and other agents pause while an agent performs inference. Such protocols obscure a fundamental challenge of real-time constraints, where the world continues to evolve and delayed decisions may become outdated before execution. Motivated by this limitation, we introduce ROAR, a multiplayer action-shooter game environment for evaluating LMM-based agents in multi-agent settings under real-time constraints. ROAR covers cooperative, competitive, mixed cooperative-competitive, and basic single-agent scenarios. We further propose a Multi-Agent Asynchronous Gym interface that decouples environment evolution from agent reasoning by running the environment and each agent in separate threads, enabling the environment to advance at fixed time intervals without waiting for LMM inference and allowing agents to reason independently without blocking one another. ROAR also features convenient deployment, multimodal observations, and customizable levels. ROAR offers a practical testbed for advancing LMM-based agents beyond static, turn-based evaluation toward real-time multi-agent decision-making.
\end{abstract}

\section{Introduction}

The rapid advancement of large multimodal models (LMMs) has been transformative \citep{team2023gemini, liu2023visual, yang2025qwen3, li2026toward}. By integrating textual, visual, and auditory perception \citep{he2024multi, zhu2024llava, liu2026palm, yu2025core3d} with reasoning \citep{wei2022chain, liu2026tracing, shen2026fine}, long-horizon planning \citep{yao2022react, shen2026decoding}, and dynamic decision-making \citep{wang2026towermind}, LMMs are increasingly becoming central to a new agentic paradigm in which they function as agents that perceive the environment and take actions accordingly \citep{bolton2025sima, park2023generative, liu2023agentbench, zhang2025metaagent}. This paradigm has gained traction in multi-agent settings \citep{li2024survey, li2025embodied}, including robot teaming \citep{kang2025viki, song2025collabot}, social interaction scenarios \citep{yang2024v, liu2026lmagent}, and autonomous vehicle fleets \citep{chiu2025v2v}, where agent interactions can be cooperative, competitive, or mixed \citep{wong2023deep}.

Despite growing interest in LMM-based agents in multi-agent settings,  existing benchmarks fail to capture the \textit{real-time constraints} inherent in these settings. Most existing benchmarks, such as LLM-PySC2 \citep{li2024llm}, TeamCraft \citep{long2024teamcraft}, and TDW-Cook \citep{zhang2024combo}, are fundamentally \textit{turn-based}: both environmental dynamics and other agents’ actions pause during LMM inference, allowing agents unconstrained time for reasoning and action generation. This assumption is inconsistent with realistic real-time multi-agent scenarios, where the environment and other agents continue to evolve independently of the agent’s reasoning, making delayed decisions potentially obsolete upon execution. As a result, real-time constraints in multi-agent settings require LMM-based agents to maintain both strong reasoning quality and timely responsiveness \citep{wen2025real, kang2025win, feng2025}. However, benchmarks for evaluating LMM-based agents under real-time constraints in multi-agent settings remain lacking.

\begin{figure}[ht]
    \centering
    \includegraphics[width=1.0\linewidth]{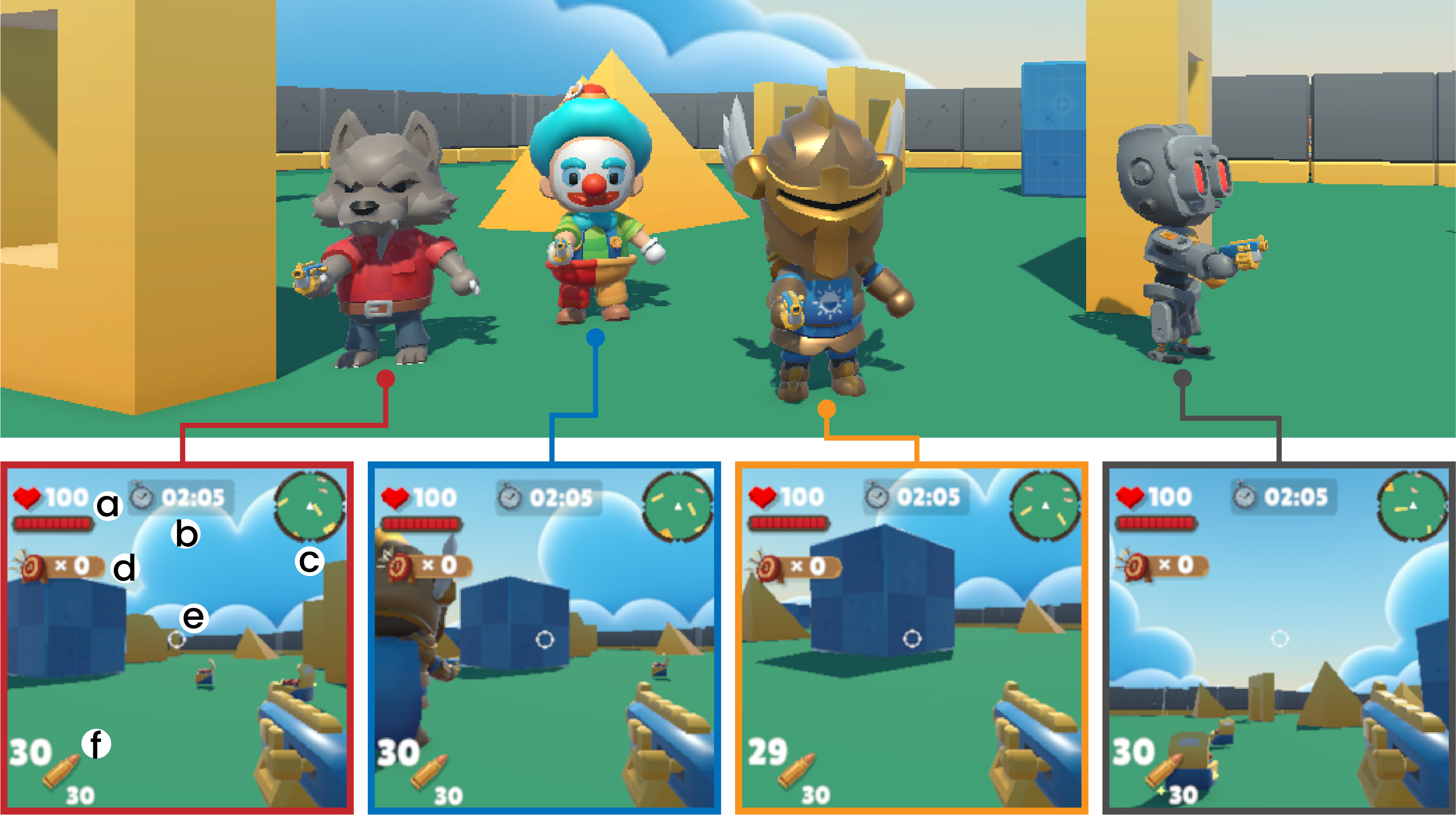}
    \caption{ROAR is a multiplayer action-shooter game environment. Top: A third-person view of the agents and the environment in ROAR. Bottom row: First-person views of the four agents shown above during gameplay, with each image corresponding to one agent. Labels (a)–(f) denote the agent’s current health, remaining episode time, minimap, number of defeated opponents, crosshair, and current/maximum ammunition, respectively.}
    \label{fig:screenshot}
\end{figure}


In this paper, we propose \textbf{ROAR} (a\textbf{R}ena f\textbf{O}r multi-\textbf{A}gent \textbf{R}eal-time challenges), a multiplayer action-shooter game environment for evaluating the performance of LMM-based agents in multi-agent settings under real-time constraints. A screenshot is provided in Figure~\ref{fig:screenshot}. In this environment, agents must navigate complex obstacle-filled arenas and win by collecting coins or shooting opponents. ROAR adopts our customized \textbf{Multi-Agent Asynchronous Gym} interface, in which the environment and each agent run in separate threads, allowing the environment to continue advancing at a fixed timestep even while the agents are reasoning, and ensuring that agents do not block one another. It supports cooperative, competitive, and mixed cooperative-competitive multi-agent settings \cite{zhang2021multi}, which broadly characterize the main categories of multi-agent interaction. Moreover, these settings evaluate LMM-based agents from complementary perspectives. The cooperative setting assesses whether agents can coordinate toward shared goals, while the competitive setting tests whether they can adapt against strategic opponents. The mixed setting further probes whether agents can balance team-level coordination with opponent-aware decision-making.


The choice of an action-shooter game in this work is motivated by its characteristics, which naturally impose demands on both reasoning quality and response speed: (1) Action-shooter games are mechanically streamlined yet far from trivial, posing considerable challenges to the multimodal perception and reasoning abilities of LMMs \cite{kempka2016vizdoom, jaderberg2019human, fu2025learning, yu2026elsa3d}. While the action space consists only of movement, aiming, and shooting, agents are required to extract informative cues from their perception of a 3-dimensional (3D) environment, perform situation-aware reasoning, and formulate targeted strategies to accomplish objectives such as collecting items or defeating opponents. (2) Action-shooter games impose strict real-time constraints because their core interactions are highly time-sensitive and adversarial. Opponents continue to move, aim, and attack during the agent’s reasoning process, while opportunities for targeting, shooting, or evasion may last only briefly. As a result, delayed decisions can quickly become outdated, increasing the likelihood of losing targets or taking hits \cite{de2024will}. Appendix~\ref{appendix:motivation_exp} presents a motivating experiment highlighting action-shooter games' time sensitivity.

ROAR also provides additional features to support further research: (1) ROAR supports split-screen multiplayer within a single executable, without requiring multiple game instances or network connections. Moreover, ROAR is lightweight, requiring only 0.2 GB of disk space and memory, and can perform CPU-based rendering without a dedicated GPU. These properties simplify deployment and improve the convenience of using ROAR for research \cite{stodden2014implementing}. (2) ROAR incorporates a multimodal observation space that includes pixels, depth, text, game states, and audio, enabling utilization and evaluation of the multimodal perception and reasoning capabilities of LMMs \cite{bie2025omniplay}. (3) ROAR provides a map editor for manual editing and procedural generation, enabling task-specific map customization while reducing the risk of data contamination \cite{jacovi2023stop, white2024livebench, li2025metawild}, details are provided in Appendix~\ref{appendix:env_custom}.


The main contributions of this work are as follows: (1) We propose ROAR, a multiplayer action-shooter environment, together with a Multi-Agent Asynchronous Gym interface, with the aim of encouraging the community to move beyond static, turn-based evaluation toward dynamic real-time decision-making. (2) We empirically find that real-time constraints affect LMM-based agents differently across multi-agent settings. They degrade all evaluated models in cooperative settings, while in competitive and mixed settings, the degradation of stronger models can narrow the observed performance gap among agents. This suggests that future research on real-time multi-agent systems should treat latency as a system-level factor, rather than focusing solely on optimizing individual agents. (3) We further investigate whether centralized control and agentic frameworks can improve multi-agent cooperation under real-time constraints, and reveal that their coordination and reasoning benefits can be offset by the additional latency they introduce. This underscores a fundamental paradigm shift for future research: in real-time environments, the value of a decision is strictly bounded by its timeliness, suggesting that future real-time agents should be evaluated and designed under a joint objective of reasoning quality, coordination, and action timeliness.



\begin{table}[t]
\centering
\caption{Comparison of ROAR with other benchmarks for LMM-based agents in multi-agent settings. \textbf{Real-Time} indicates whether the environment supports real-time constraints. \textbf{Player Mode} denotes the agent setting, where \textbf{Coop.}, \textbf{Comp.}, \textbf{Mixed}, and \textbf{Single} represent the multi-agent cooperative, multi-agent competitive, multi-agent mixed cooperative-competitive, and single-agent settings, respectively. \textbf{Observation} denotes the observation modalities provided by the environment, where \textbf{State} corresponds to structured game state observations. \halfcheck denotes partial real-time support; MineLand uses non-blocking, time-driven execution with New/Resume gating, making it neither strictly turn-based nor fully real-time.}
\label{tab:compare_envs}
\setlength{\tabcolsep}{0.6mm}
\begin{tabular}{l|c|cccc|ccccc}
\toprule
\multirow{2}{*}{\makecell{\textbf{Environment} }} &\multirow{2}{*}{\makecell{\textbf{Real-time} }} & \multicolumn{4}{c|}{\makecell{\textbf{Play Mode} }} & \multicolumn{5}{c}{\makecell{\textbf{Observation} }} \\
\cmidrule(lr){3-6}\cmidrule(lr){7-11}
 && \textbf{Coop.} & \textbf{Comp.} & \textbf{Mixed}& \textbf{Single} & \textbf{Pixel}& \textbf{Depth}& \textbf{Text}& \makecell{\textbf{State} }& \textbf{Audio} \\
\midrule
LLM-PySC2 \citep{li2024llm} & \xmark & \xmark & \cmark& \xmark& \xmark& \cmark& \xmark& \cmark& \xmark& \xmark \\
TeamCraft \citep{long2024teamcraft} & \xmark & \cmark & \xmark& \xmark& \xmark& \cmark& \xmark& \xmark& \cmark& \xmark \\
MineLand \citep{yu2024mineland}& \halfcheck& \cmark& \cmark& \xmark& \cmark& \cmark& \xmark& \xmark& \xmark& \cmark\\
WhodunitBench \citep{xie2024whodunitbench}& \xmark& \xmark& \xmark& \cmark& \xmark& \cmark& \xmark& \cmark& \xmark& \xmark\\
MAP-THOR \citep{nayak2024long}& \xmark& \cmark& \xmark& \xmark& \cmark& \cmark& \xmark& \cmark& \cmark& \xmark\\
TDW-Cook \citep{zhang2024combo}& \xmark& \cmark& \xmark& \xmark& \xmark& \cmark& \cmark& \xmark& \cmark& \xmark\\
VIKI-Bench \citep{kang2025viki}& \xmark& \cmark& \xmark& \xmark& \xmark& \cmark& \xmark& \cmark& \xmark& \xmark\\
MARS-Bench \citep{wang2026mars}& \xmark& \cmark& \xmark& \xmark& \xmark& \cmark& \cmark& \xmark& \xmark& \xmark\\
\midrule
\textbf{ROAR (Ours)}& \cmark& \cmark& \cmark& \cmark& \cmark& \cmark& \cmark& \cmark& \cmark& \cmark\\
\bottomrule
\end{tabular}
\end{table}

\section{Related Work}


\textbf{Real-Time Constraints for Intelligent Agents}. Real-time constraints have become increasingly important for intelligent agents in interactive applications. Prior work has studied this issue in reinforcement learning by extending MDPs to delayed or time-constrained settings \cite{ramstedt2019real, chen2021delay, derman2021acting, wikman2025adaptive, bellman1957markovian}, and in large language models research through benchmarks and methods for real-time reasoning \cite{kang2025win, wen2025real, schipper2025pillagerbench, yi2026multi}. Recent LMM-based agent benchmarks further evaluate real-time decision-making in video game environments \cite{lu2025cultivating, park2025orak, zhang2025videogamebench, ouyang2026gameworld, tan2025stardojo}. However, these works mainly focus on single-agent settings, leaving real-time constraints for LMM-based agents in multi-agent settings underexplored.


\textbf{LMM-Based Agents in Multi-Agent Settings}. Recent advances in LMMs have enabled agents that integrate multimodal perception with reasoning, planning, and action execution \cite{xie2024large}, leading to applications in GUI assistance \cite{wang2024mobile, zhang2025appagent}, game development \cite{tang2024autogameui, yang202590, zhang2025can}, and multimodal navigation \cite{paul2022avlen, qiao2025navbench}. LMM-based agents have also been explored in multi-agent domains, including embodied cooperation \cite{kang2025viki}, collaborative manipulation \cite{song2025collabot}, social simulation \cite{liu2026lmagent}, and cooperative autonomous driving \cite{chiu2025v2v}. However, existing multi-agent benchmarks such as LLM-PySC2 \cite{li2024llm}, TeamCraft \cite{long2024teamcraft}, TDW-Cook \cite{zhang2024combo}, and VIKI-Bench \cite{kang2025viki} largely overlook real-time constraints.

\textbf{Video Game--Based Environments and Benchmarks}. Video games provide controllable, scalable, and safe testbeds for studying intelligent agents in interactive settings \cite{sweetser2024large}. Environments span genres, including arcade games \cite{bellemare2013arcade, pleines2025pokemon}, open-world games \cite{fan2022minedojo, albrecht2022avalon}, real-time strategy games \cite{vinyals2017starcraft, wang2026towermind, wei2022honor}, life-simulation games \cite{carroll2019utility, gong2024mindagent, ahn-etal-2025-flashadventure}, sports games \cite{kurach2020google, jia2020fever}, action games \citep{chen2025combatvla, tan2025lumine, yue2026scaling} and first-person shooter games \cite{kempka2016vizdoom, beattie2016deepmind}. In this work, we propose a multiplayer action-shooter environment for evaluating LMM-based agents in multi-agent settings under real-time constraints. Table~\ref{tab:compare_envs} and Appendix~\ref{appendix:fps} compare ROAR with prior LMM-agent benchmarks and shooter-game environments, respectively.

\section{ROAR Environment}

ROAR is developed using the Unity game engine, with the Universal Render Pipeline employed to enable real-time lighting and shading in 3D scenes. It interfaces with machine learning via the Unity ML-Agents Toolkit \cite{juliani2018unity} and operates through our customized Multi-Agent Asynchronous Gym interface. Its codebase is fully open-source and requires no license application or purchase.

\begin{figure}[ht]
    \centering
    \includegraphics[width=1.0\linewidth]{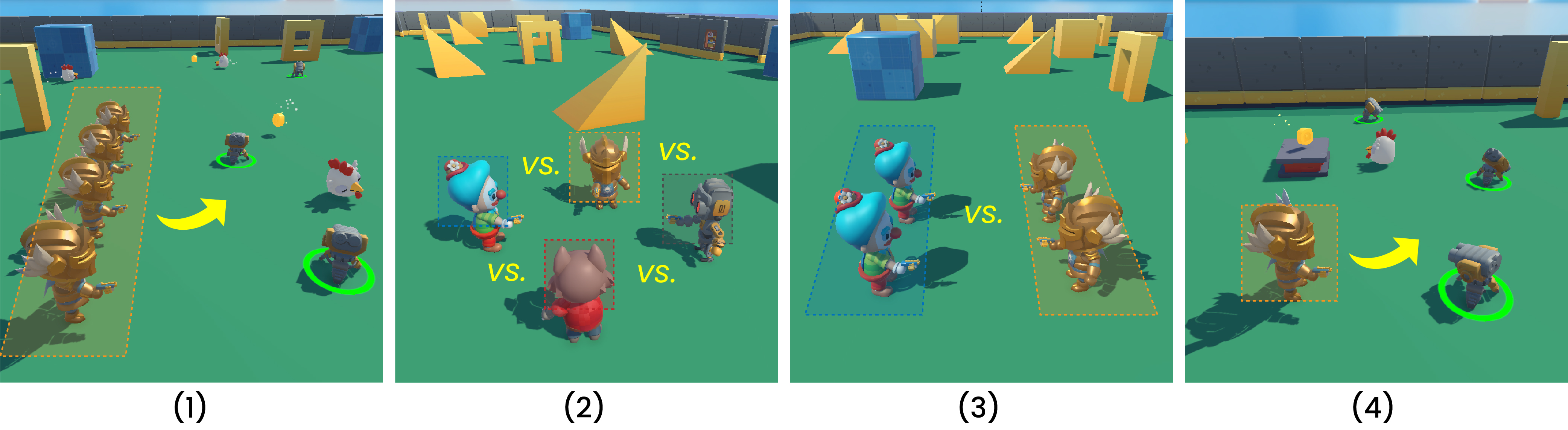}
    \caption{Illustration of the four levels in ROAR. Agents on the same team share the same 3D model and are marked with same-colour rectangles. From left to right: (1) Cooperative Level (four agents collaborate to complete the task); (2) Competitive Level (four agents compete against each other); (3) Mixed Cooperative-Competitive Level (two teams of two with intra-team cooperation and inter-team competition); and (4) Basic Single-Agent Level (a single agent completes the task).}
    \label{fig:levels}
\end{figure}

\subsection{Game Mechanics}
\label{sec:game_mech}
In ROAR, the cooperative, competitive, mixed cooperative-competitive, and basic single-agent settings are each realized as independent levels with different maps and task objectives, as illustrated in Figure~\ref{fig:levels}. Players are required to achieve task objectives such as collecting coins and defeating opponents through movement and shooting, within five minutes. See Appendix~\ref{appendix:game_details} for more details.

\textbf{Cooperative Level}: This level involves four agents whose objective is to cooperatively collect a specified number of coins scattered throughout the arena. Turrets positioned across the map attack the agents during coin collection, forcing the agents to either avoid turret fire or destroy the turrets. Non-hostile chickens provide extra team rewards when hunted. Before the time limit is reached, an episode terminates if the team collects the required number of coins or all agents are eliminated.

\textbf{Competitive Level}: This level contains four agents whose objective is to become the last survivor. After the episode begins, a toxic zone emerges from the perimeter of the arena, causing any agent within it to lose health rapidly. The zone gradually contracts toward the center and eventually covers the entire arena within five minutes, progressively intensifying adversarial interactions among agents. The episode terminates immediately if only one agent remains alive before the time limit is reached.

\textbf{Mixed Cooperative-Competitive Level}: This level contains four agents, divided into two teams of two. The objective is to eliminate the opposing team while ensuring that at least one teammate survives. This level also features a toxic zone, following the same rules as in the competitive level. The episode ends before the time limit is reached when all agents on either team have been eliminated.

\textbf{Basic Single-Agent Level}: This level mirrors the cooperative level but with only one agent.

\begin{figure}[ht]
    \centering
    \includegraphics[width=1.0\linewidth]{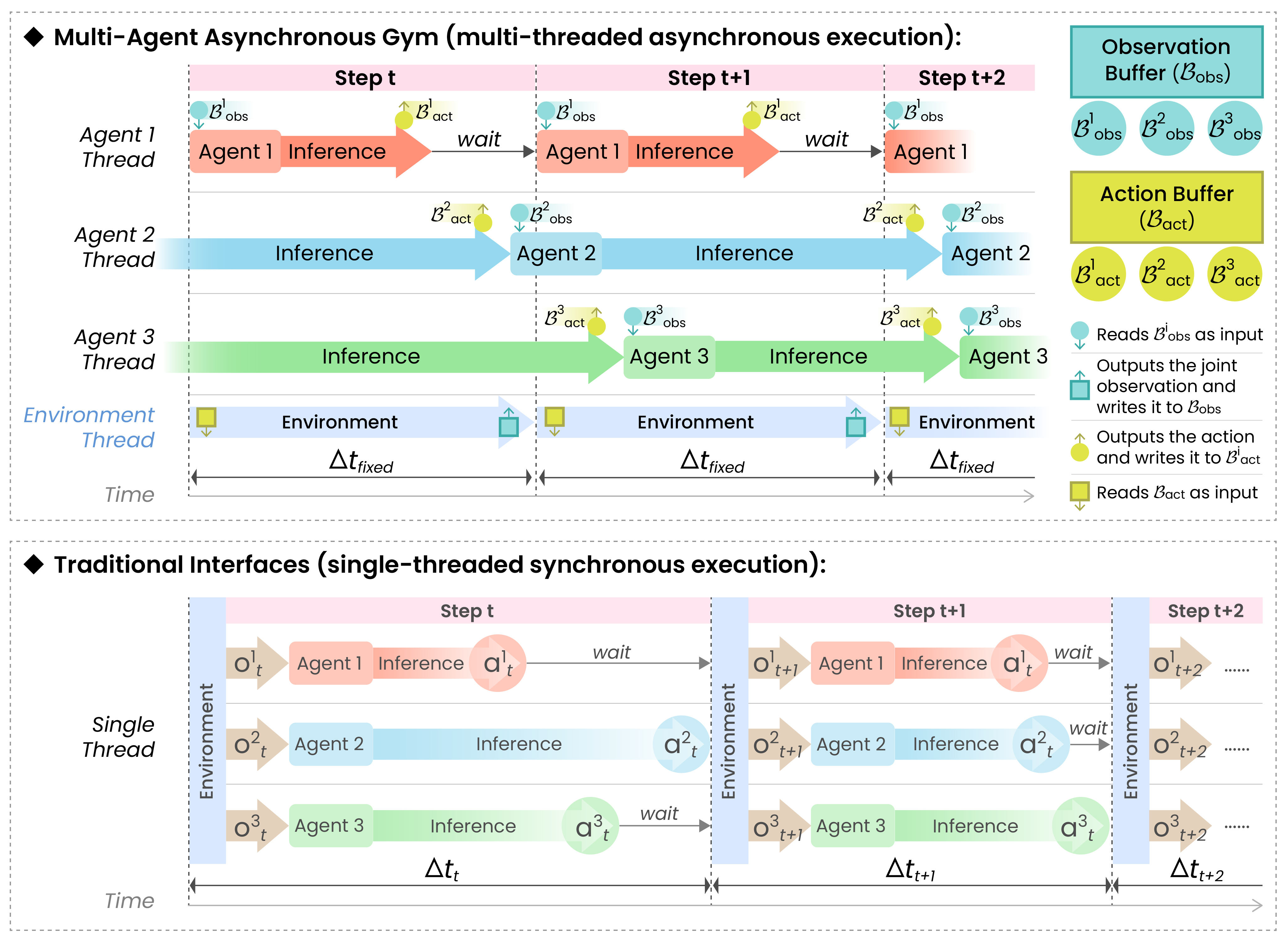}
    \caption{Comparison between our Multi-Agent Asynchronous Gym interface and traditional interfaces. Top: In Multi-Agent Asynchronous Gym, the environment and each agent run on separate threads. The environment advances at a fixed interval $\Delta t_{\mathrm{fixed}}$ without waiting for agent inference, while each agent independently reads its latest observation from the observation buffer $\mathcal{B}{\mathrm{obs}}^i$ and writes its generated action to the action buffer $\mathcal{B}{\mathrm{act}}^i$. Bottom: Traditional interfaces execute all agents and the environment within a single thread, where the environment must wait for all agents to finish inference before proceeding to the next step, resulting in variable step durations $\Delta t_t$.}
    \label{fig:framework}
\end{figure}

\subsection{Multi-Agent Asynchronous Gym}
Traditional interfaces (e.g., OpenAI Gym \cite{brockman2016openai}) are primarily designed for turn-based interaction, where the environment and all agents are executed within a single thread under a synchronous stepping scheme. The environment waits for all agents to complete their inference, forcing faster agents to idle while slower ones finish. Such a synchronization mechanism fails to reflect multi-agent real-time settings, where the environment and other agents evolve continuously during an agent’s reasoning process. We therefore introduce a custom Multi-Agent Asynchronous Gym interface, in which the environment and each agent run on dedicated threads, such that environment evolution and the reasoning processes of all agents are fully decoupled and proceed without mutual blocking or waiting. The interface is applicable beyond ROAR and can be readily integrated into environments based on traditional interfaces to equip them with real-time constraints.

We formalize the framework as a tuple $( \mathcal{N}, \mathcal{S}, \mathcal{O}, \mathcal{A}, \mathcal{T}, \mathcal{B}_{\mathrm{obs}}, \mathcal{B}_{\mathrm{act}}, \bar{a}, \phi, \Delta t_{\mathrm{fixed}}, r, T )$. $\mathcal{N} = \{1, \ldots, n\}$ denotes the finite set of $n$ agents. $\mathcal{S}$ denotes the finite set of states. $\mathcal{O} = \{\mathcal{O}^i \mid i \in \mathcal{N}\}$ denotes the finite set of observation spaces, where $\mathcal{O}^i$ is the observation space of agent $i$, joint observation $\boldsymbol{o} \in \mathcal{O} = \mathcal{O}^1 \times \cdots \times \mathcal{O}^n$. $\mathcal{A} = \{\mathcal{A}^i \mid i \in \mathcal{N}\}$ denotes the finite set of action spaces, where $\mathcal{A}^i$ is the action space of agent $i$, joint action $\boldsymbol{a} \in \mathcal{A} = \mathcal{A}^1 \times \cdots \times \mathcal{A}^n$. $\mathcal{T} = \{\tau_{\mathrm{env}}\} \cup \{\tau^i_{\mathrm{agent}} \mid i \in \mathcal{N}\}$ denotes the finite set of threads, consisting of the environment thread $\tau_{\mathrm{env}}$, in which the environment evolves, and the set of agent threads $\{\tau^i_{\mathrm{agent}} \mid i \in \mathcal{N}\}$, in which the agents perform decision-making. $\mathcal{B}_{\mathrm{obs}} = \{\mathcal{B}_{\mathrm{obs}}^i \mid i \in \mathcal{N}\}$ denotes the finite set of observation buffers storing the latest observations (for inter-thread synchronization), where $\mathcal{B}_{\mathrm{obs}}^i$ stores the latest observation of agent $i$. $\mathcal{B}_{\mathrm{act}} = \{\mathcal{B}_{\mathrm{act}}^i \mid i \in \mathcal{N}\}$ denotes the finite set of action buffers storing the latest actions to be executed by the agents (for inter-thread synchronization), where $\mathcal{B}_{\mathrm{act}}^i$ stores the latest action to be executed by agent $i$. $\bar{a} = \{\bar{a}^i \mid i \in \mathcal{N}\}$ denotes the finite set of default actions of the agents, where $\bar{a}^i$ is the default action of agent $i$. $\phi$ denotes the dynamics model, where $\phi(s', \boldsymbol{o}' \mid s, \boldsymbol{a}) \in [0, 1]$ specifies the probability of transitioning to state $s'$ and receiving joint observation $\boldsymbol{o}'$ when the agents take joint action $\boldsymbol{a}$ in state $s$. $\Delta t_{\mathrm{fixed}}$ denotes the fixed time interval between timesteps at which the environment performs state transitions. $r$ denotes the reward function, where $r(s, \boldsymbol{a})$ specifies the reward received when the agents take joint action $\boldsymbol{a}$ in state $s$. $T$ denotes the episode time horizon, with timesteps indexed by $t = 0, \dots, T$.

The workflow of the Multi-Agent Asynchronous Gym is illustrated in Figure~\ref{fig:framework}: Each agent retrieves the latest observation from its observation buffer $\mathcal{B}_{\mathrm{obs}}^i$ and then performs reasoning and decision-making. Once the decision process is completed, the agent writes the resulting action to be executed into its action buffer $\mathcal{B}_{\mathrm{act}}^i$, then retrieves the latest observation from $\mathcal{B}_{\mathrm{obs}}^i$ and begins the next round of reasoning. Meanwhile, the environment advances continuously at a fixed interval $\Delta t_{\mathrm{fixed}}$. At each timestep $t$, the environment constructs the joint action $\boldsymbol{a}_t$ by reading from $\mathcal{B}_{\mathrm{act}}$: if $\mathcal{B}_{\mathrm{act}}^i$ contains a valid action, that action is used as the $i$-th component of $\boldsymbol{a}_t$ and $\mathcal{B}_{\mathrm{act}}^i$ is then reset to an invalid value; otherwise, the default action $\bar{a}^i$ is used. After the joint action $\boldsymbol{a}_t$ is formed, the environment performs one state transition according to $\phi$, obtains the next joint observation $\boldsymbol{o}_{t+1}$, and stores the corresponding observations in $\mathcal{B}_{\mathrm{obs}}^i$. The corresponding pseudocode is provided in Appendix~\ref{appendix:pseudocode}.

\subsection{Environment Interface}

\textbf{Observations}. The observation space of each agent $\mathcal{O}^i$ consists of five distinct modalities, each reflecting the agent’s perception of the game environment at the current timestep in a different form. (1) The pixel observation is an RGB image of the raw game screen from each agent’s first-person perspective, with a default resolution of 512 × 512 that can be adjusted according to research needs. (2) The depth observation provides the per-pixel depth values of the raw game screen from each agent’s perspective, with the same resolution as the pixel observation. (3) The text observation represents the internal game state in JSON format and is divided into two parts: a shared global state and agent-specific states. For example, \texttt{RemainingTime} denotes the shared remaining time in the current episode, while \texttt{CurrentAmmo} specifies the current ammunition amount of each agent. (4) The game state observation is a one-dimensional array obtained by flattening the aforementioned JSON-formatted data. (5) The audio observation consists of an array of all sound sources audible to each agent. Each sound-source entry contains a set of attributes, such as its tag, direction, and distance. Researchers may use the audio observation directly in this structured form, or alternatively employ our built-in audio mixing tool to combine all audible sound sources into a raw audio signal for the agent. Examples of each observation modality are provided in Appendix~\ref{appendix:obs}.

\begin{figure}[ht]
    \centering
    \includegraphics[width=1.0\linewidth]{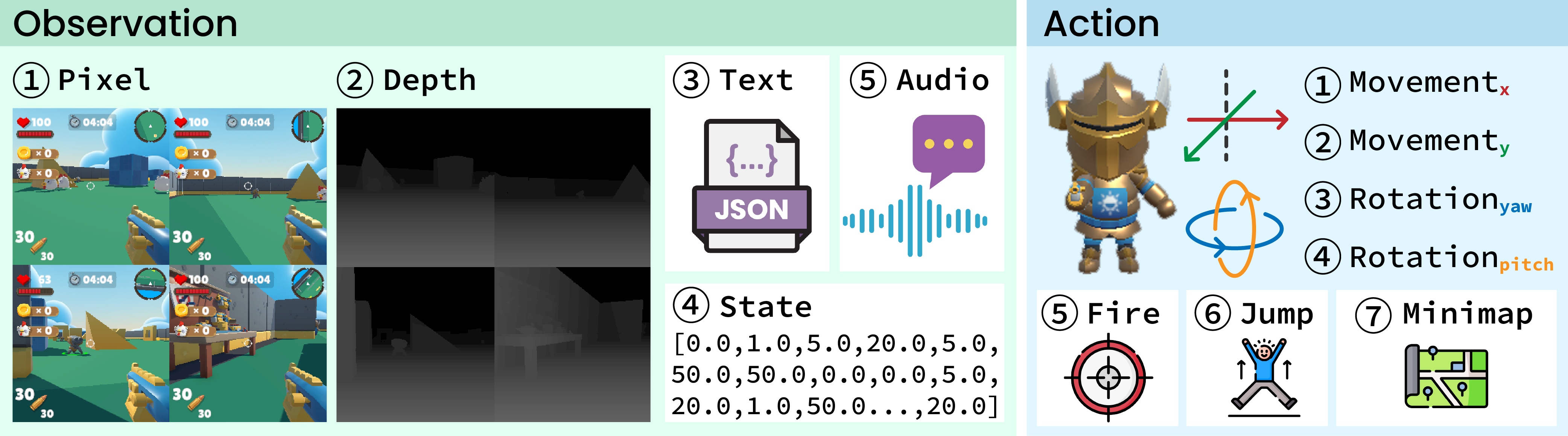}
    \caption{The ROAR environment provides a multimodal observation space, including pixels, depth, text, game states, and audio. The action space is a seven-dimensional vector representing horizontal (left–right) and vertical (forward–backward) movement, yaw (left–right) and pitch (up–down) camera rotations, as well as firing, jumping, and minimap usage.}
    \label{fig:interface}
\end{figure}

\textbf{Actions}. The action space for each agent $\mathcal{A}^i$ in this environment is represented as a seven-dimensional vector, where each action $a$ is defined as: \[ a = (movement_{\mathrm{x}}, movement_{\mathrm{y}}, rotation_{\mathrm{yaw}}, rotation_{\mathrm{pitch}}, fire, jump, minimap). \] Specifically, $movement_{\mathrm{x}}$ and $movement_{\mathrm{y}}$ denote movement along the local horizontal (left-right) and vertical (forward-backward) axes, respectively; $rotation_{\mathrm{yaw}}$ and $rotation_{\mathrm{pitch}}$ denote rotations in the yaw (left–right) and pitch (up–down) directions, respectively; $fire$, $jump$, and $minimap$ denote firing, jumping, and toggling the enlarged minimap display, respectively. See Appendix~\ref{appendix:action}.

\textbf{Reward}.
\label{sec:reward}
ROAR adopts sparse rewards, with each agent receiving a final reward (episode score) ranging from 0 to 1 based on level-specific rules. In cooperative and basic single-agent levels, the score reflects coin collection and chicken hunting; in competitive and mixed levels, it reflects opponents defeated and survival time. See Appendix~\ref{appendix:reward} for level-specific reward formulations.


\section{Experiment}

\subsection{Experiment Setup}
\label{sec:exp_setup}

\textbf{Models.} We evaluate five LMMs: three proprietary models, GPT-5.4 \cite{OpenAI}, Gemini-3.1 Pro \cite{team2023gemini}, and Claude Haiku-4.5 \cite{Anthropic1}; and two open-source models, Qwen-3.5 9B \cite{yang2025qwen3} and Gemma-4 E4B \cite{kamath2025gemma}. Proprietary models are accessed through their official APIs, whereas open-source models are deployed locally on an NVIDIA H200 GPU. For all models, we report end-to-end latency, defined as the wall-clock time between issuing a request and receiving the complete model response. We evaluate LMMs using a zero-shot prompting strategy, with the thinking mode disabled for all models. Each model receives both text and pixel observations as inputs. Multi-agent control is decentralized: each agent controls one game character and makes decisions based solely on its limited local observations, with no inter-agent communication. In this experiment, agents are directly controlled by LMMs without the use of any agentic frameworks. Additional experimental details are provided in Appendix~\ref{appendix:exp}.

\textbf{Metrics.} We use the \textbf{episode score} described in the reward description in Section~\ref{sec:reward} as our evaluation metric, with a value range of $[0, 1]$. For the cooperative level and basic single-agent level, we run 20 random seeds for agents controlled by each model and report the average episode score. For the competitive level, we evaluate through a full combinatorial tournament. Each match includes four distinct models, yielding $\binom{5}{4}=5$ unique lineups. To mitigate confounding effects from player-character 3D models and spawn positions, we evaluate each lineup under four cyclic permutations of player assignments, each with five random seeds. We then report the average episode score across all episodes in which each model participates. For the mixed cooperative-competitive level, we follow the same protocol as in the competitive level, with two distinct models per match, yielding $\binom{5}{2}=10$ unique lineups. Each lineup is evaluated with swapped player assignments, with five random seeds for each assignment. We report each model’s average episode score across all participating episodes. In addition to the episode score, we report two in-game statistics: \textbf{shooting accuracy} and \textbf{incoming hit rate}. Shooting accuracy is computed as
\(\mathrm{Acc}_{\mathrm{shoot}} = N_{\mathrm{hit}} / N_{\mathrm{shot}}\),
where \(N_{\mathrm{hit}}\) is the number of shots hitting valid targets (i.e., chickens, turrets, and opponent agents) and \(N_{\mathrm{shot}}\) is the total number of shots fired. Incoming hit rate is computed as
\(\mathrm{Rate}_{\mathrm{incoming}} = N_{\mathrm{turret\_hit}} / N_{\mathrm{turret\_lock}}\),
where \(N_{\mathrm{turret\_hit}}\) is the number of times an agent is hit by turrets and \(N_{\mathrm{turret\_lock}}\) is the number of times the agent is locked onto by turrets. These two metrics directly reflect an agent’s ability to hit targets and avoid incoming attacks, respectively. More details are provided in the Appendix~\ref{appendix:exp_1}.


\textbf{Environment Stepping.} Our experiments evaluate the performance of LMM-based agents in both the Multi-Agent Asynchronous Gym with real-time constraints and the traditional turn-based interface. For experiments conducted with the Multi-Agent Asynchronous Gym, we consider two fixed environment step intervals, $\Delta t_{\mathrm{fixed}} = 40$ ms (25 Hz) and $\Delta t_{\mathrm{fixed}} = 250$ ms (5 Hz). The former reflects a commonly used update frequency in robotic systems \cite{song2026fast, ashtiani2021hybrid}, whereas the latter approximates human simple reaction time, defined as the time required to respond as quickly as possible to a stimulus \cite{woods2015factors}. The default action for all agents in the experiments is $[0,0,0,0,0,0,0]$, which corresponds to taking no action, including no movement, camera rotation, firing, jumping, or opening the minimap (see Appendix~\ref{appendix:action}). In experiments with the traditional turn-based interface, the environment waits for all agents to finish inference at each step before advancing to the next step.

\begin{table*}[t]
\setlength{\tabcolsep}{1.0mm}
    \centering
     \caption{Performance of different LMMs under the traditional turn-based interface and our Multi-Agent Asynchronous Gym interface with fixed time intervals of 250 ms and 40 ms. \textbf{Latency} denotes the end-to-end model latency, reported using P50 and P90. \textbf{Coop.}, \textbf{Comp.}, \textbf{Mixed}, and \textbf{Single} report the episode scores on the cooperative, competitive, mixed cooperative-competitive, and basic single-agent levels, respectively. \textbf{Avg.} reports the average episode score across the four levels. \textbf{Shooting} and \textbf{Incoming} denote shooting accuracy and incoming hit rate, respectively. Bold numbers indicate the highest value in each column. Tables with standard errors are provided in the Appendix~\ref{appendix:exp_res}}
    \label{tab:exp_result}
    \setlength{\tabcolsep}{0.6mm}
    \begin{tabular}{c l| c c c c >{\color{gray!80}}c c c c}
        \toprule
        \textbf{Model} & 
        \textbf{Latency (ms)} &
        $\boldsymbol{\Delta t_{\mathrm{fixed}}}$ \textbf{(ms)} &
        \textbf{Coop.} & 
        \textbf{Comp.} & 
        \textbf{Mixed} & 
        \textbf{Single} & 
        \textbf{Avg.} &
        \textbf{Shooting} &
        \textbf{Incoming} \\
        
        \midrule
        \multirow{3}{*}{\textbf{GPT-5.4}}
& \multirow{3}{*}{\makecell{P50: 1458.5\\P90: 1946.1}} 
& Turn-based & \textbf{0.31} & \textbf{0.54} & \textbf{0.65} & \textbf{0.38} & \textbf{0.47} & \textbf{0.54} & 0.26 \\
&  
& 250         & $0.07$          & $0.36$          & $0.43$          & $0.16$          & 0.26 & 0.29 & 0.71\\
&  
& 40        & $0.02$ & 0.33 & 0.30 &  $0.12$ & 0.19 & 0.07 & \textbf{0.88} \\
\midrule
        \multirow{3}{*}{\textbf{\makecell{Gemini-3.1\\Pro}}}
& \multirow{3}{*}{\makecell{P50: 3235.2\\P90: 3995.3}} 
& Turn-based &  \textbf{0.78} & \textbf{0.82} & \textbf{0.85} &  \textbf{0.93} &  \textbf{0.87} &  \textbf{0.82} & 0.31 \\
&  
& 250         & $0.13$          & 0.68          & 0.71          & $0.22$          & 0.44 & 0.61 & 0.66\\
&  
& 40        & $0.04$ & 0.54 & 0.63 & $0.16$ & 0.34 & 0.23 &  \textbf{0.89} \\
        
        \midrule

        \multirow{3}{*}{\textbf{\makecell{Claude\\Haiku 4.5}}}
& \multirow{3}{*}{\makecell{P50: 1861.0\\P90: 2352.6}} 
& Turn-based & \textbf{0.05} & 0.18 & 0.20 & \textbf{0.08} & \textbf{0.13} & 0.00 & 0.51 \\
&  
& 250         & $0.00$          & \textbf{0.21}          & 0.21         & $0.04$          & 0.12 & 0.00 & 0.87 \\
&  
& 40        & $0.00$ & \textbf{0.21} & \textbf{0.22} & $0.00$ & 0.11 & 0.00 & \textbf{0.88} \\
        
        \midrule

        \multirow{3}{*}{\textbf{\makecell{Qwen-3.5\\9B}}}
& \multirow{3}{*}{\makecell[l]{P50: 96.6\\P90: 226.5}} 
& Turn-based & \textbf{0.06} & 0.09 & 0.13 & \textbf{0.12} & \textbf{0.10} & \textbf{0.04} & 0.57 \\
&  
& 250         & $0.00$          & 0.11          & \textbf{0.15}          & $0.01$          & 0.07 & 0.03 & 0.61 \\
&  
& 40        & $0.00$ & \textbf{0.12} & \textbf{0.15} & $0.00$ & 0.07 & 0.03 & \textbf{0.71} \\
        
        \midrule

        \multirow{3}{*}{\textbf{\makecell{Gemma-4\\E4B}}}
& \multirow{3}{*}{\makecell[l]{P50: 64.5\\P90: 136.8}} 
& Turn-based & \textbf{0.05} & 0.07  & 0.13 & \textbf{0.06} & \textbf{0.08} & 0.00 & 0.67 \\
&  
& 250         & $0.02$          & 0.10          & 0.15          & $0.00$          & 0.07 & 0.00 & 0.69 \\
&  
& 40        & $0.00$ & \textbf{0.11} & \textbf{0.16} & $0.00$ & 0.07 & 0.00 & \textbf{0.75} \\
        
        \bottomrule
    \end{tabular}
   
\end{table*}


\subsection{Experiment Results}
The results in Table~\ref{tab:exp_result} show that real-time constraints disrupt the reasoning-to-action pipeline of LMM-based agents. Overall, all evaluated models exhibit lower average performance across the four levels under real-time constraints, while the impact varies across different multi-agent settings.

\textbf{Performance in the Cooperative Level.} In this level, real-time execution degrades the performance of all evaluated models and leads to the largest performance drop among the four levels. For example, Gemini-3.1 Pro decreases from 0.78 in the turn-based setting to 0.04 at 40 ms. These results indicate that cooperative multi-agent tasks are particularly sensitive to delayed decisions: agents must not only reason about their own observations, but also maintain timely coordination with teammates while the environment continues to evolve. 

\textbf{Performance in the Competitive and Mixed Levels.} These two levels exhibit a different pattern. For example, GPT-5.4 drops from 0.54 to 0.33 in the competitive level when moving from the turn-based interface to the 40 ms real-time setting, and from 0.65 to 0.30 in the mixed level. In contrast, Qwen-3.5 9B and Gemma-4 E4B maintain or even slightly improve their competitive and mixed scores under real-time execution, despite their weaker turn-based performance. This does not necessarily indicate that these smaller models become better real-time decision makers in an absolute sense. Rather, in competitive and mixed settings, the degradation of stronger opponents can reduce their advantage and narrow the observed performance gap. This phenomenon highlights a distinctive property of real-time multi-agent evaluation: latency can affect not only an agent's own interaction with the environment, but also the relative advantage structure among agents.

\textbf{In-Game Statistics.} Shooting accuracy and incoming hit rate provide intuitive in-game evidence of how real-time constraints disrupt the interaction loop between LMM-based agents and the environment. As the fixed time interval decreases, shooting accuracy drops while incoming hit rate increases. For example, Gemini-3.1 Pro's shooting accuracy decreases from 0.82 to 0.23, while its incoming hit rate increases from 0.31 to 0.89. This suggests that delayed decisions under real-time constraints often lead to outdated actions: when an agent fires, the target may have already moved away; when an agent attempts to evade, it may have already been hit.

\begin{table*}[t]
\setlength{\tabcolsep}{1.0mm}
    \centering
    \caption{Performance of different open-source LMMs and calibration baselines under the traditional turn-based interface and our Multi-Agent Asynchronous Gym interface with fixed time intervals of 250 ms and 40 ms. \textbf{Latency} denotes the end-to-end model latency, reported using P50 and P90. \textbf{Coop.}, \textbf{Comp.}, \textbf{Mixed}, and \textbf{Single} report the episode scores on the cooperative, competitive, mixed cooperative-competitive, and basic single-agent levels, respectively. \textbf{Avg.} reports the average episode score across the four levels. \textbf{Shooting} and \textbf{Incoming} denote shooting accuracy and incoming hit rate, respectively. Bold numbers indicate the highest value in each column. Tables with standard errors are provided in Appendix~\ref{appendix:exp_res}.}
    \label{tab:exp_result_open}
    \setlength{\tabcolsep}{0.6mm}

    \begin{tabular}{c l| c c c c >{\color{gray!80}}c c c c}
        \toprule
        \textbf{Model} & 
        \textbf{Latency (ms)} &
        $\boldsymbol{\Delta t_{\mathrm{fixed}}}$ \textbf{(ms)} &
        \textbf{Coop.} & 
        \textbf{Comp.} & 
        \textbf{Mixed} & 
        \textbf{Single} & 
        \textbf{Avg.} &
        \textbf{Shooting} &
        \textbf{Incoming} \\
        
        \midrule

        \multirow{3}{*}{\textbf{\makecell{Gemma-4\\31B}}}
        & \multirow{3}{*}{\makecell[l]{P50: 300.6\\P90: 603.9}} 
        & Turn-based 
        & \textbf{0.24} 
        & \textbf{0.43} 
        & \textbf{0.51} 
        & \textbf{0.38} 
        & \textbf{0.39} 
        & \textbf{0.32} 
        & 0.41 \\
        &
        & 250
        & 0.19
        & 0.36
        & 0.47
        & 0.31
        & 0.33
        & 0.27
        & 0.47 \\
        &
        & 40
        & 0.08
        & 0.25
        & 0.28
        & 0.13
        & 0.19
        & 0.11
        & \textbf{0.62} \\

        \midrule

        \multirow{3}{*}{\textbf{\makecell{Qwen-3.5\\27B}}}
        & \multirow{3}{*}{\makecell[l]{P50: 207.6\\P90: 478.9}} 
        & Turn-based 
        & \textbf{0.21}
        & \textbf{0.45}
        & \textbf{0.54}
        & \textbf{0.36}
        & \textbf{0.39}
        & \textbf{0.33}
        & 0.45 \\
        &
        & 250
        & 0.18
        & 0.40
        & 0.52
        & 0.31
        & 0.35
        & 0.29
        & 0.56 \\
        &
        & 40
        & 0.11
        & 0.25
        & 0.30
        & 0.15
        & 0.20
        & 0.17
        & \textbf{0.67} \\

        \midrule

        \multirow{3}{*}{\textbf{\makecell{Qwen-3.5\\9B}}}
        & \multirow{3}{*}{\makecell[l]{P50: 95.2\\P90: 215.3}} 
        & Turn-based 
        & \textbf{0.06}
        & 0.12
        & 0.13
        & \textbf{0.12}
        & \textbf{0.11}
        & \textbf{0.04}
        & 0.57 \\
        &
        & 250
        & 0.00
        & 0.13
        & \textbf{0.14}
        & 0.01
        & 0.07
        & 0.03
        & 0.61 \\
        &
        & 40
        & 0.00
        & \textbf{0.14}
        & \textbf{0.14}
        & 0.00
        & 0.07
        & 0.03
        & \textbf{0.71} \\

        \midrule

        \multirow{3}{*}{\textbf{\makecell{Gemma-4\\E4B}}}
        & \multirow{3}{*}{\makecell[l]{P50: 66.8\\P90: 134.2}} 
        & Turn-based 
        & \textbf{0.05}
        & 0.12
        & 0.12
        & \textbf{0.06}
        & \textbf{0.09}
        & 0.00
        & 0.67 \\
        &
        & 250
        & 0.02
        & 0.12
        & 0.13
        & 0.00
        & 0.07
        & 0.00
        & 0.69 \\
        &
        & 40
        & 0.00
        & \textbf{0.14}
        & \textbf{0.14}
        & 0.00
        & 0.07
        & 0.00
        & \textbf{0.75} \\

        \specialrule{1pt}{2pt}{2pt}

        \multirow{3}{*}{\textbf{\makecell{Heuristic\\Bot}}}
        & \multirow{3}{*}{\makecell[l]{P50: 0.0\\P90: 0.1}} 
        & Turn-based 
        & 0.33
        & 0.25
        & 0.27
        & 0.37
        & 0.31
        & \textbf{0.32}
        & 0.28 \\
        &
        & 250
        & \textbf{0.34}
        & 0.29
        & 0.30
        & 0.37
        & 0.33
        & 0.31
        & 0.28 \\
        &
        & 40
        & \textbf{0.34}
        & \textbf{0.31}
        & \textbf{0.32}
        & 0.37
        & \textbf{0.34}
        & \textbf{0.32}
        & 0.28 \\

        \midrule

        \multirow{3}{*}{\textbf{\makecell{Random\\Policy}}}
        & \multirow{3}{*}{\makecell[l]{P50: 0.0\\P90: 0.0}} 
        & Turn-based 
        & 0.00
        & 0.09
        & 0.10
        & 0.00
        & 0.05
        & 0.00
        & 0.85 \\
        &
        & 250
        & 0.00
        & 0.11
        & 0.11
        & 0.00
        & \textbf{0.06}
        & 0.00
        & \textbf{0.86} \\
        &
        & 40
        & 0.00
        & \textbf{0.12}
        & \textbf{0.12}
        & 0.00
        & \textbf{0.06}
        & 0.00
        & 0.85 \\

        \bottomrule
    \end{tabular}
    
\end{table*}

\subsection{Further Analysis}

\textbf{Controlled Evaluation with Locally Deployed Open-Source Models}. To further examine the consistency of the observed trends, we additionally evaluate open-source models under a unified local deployment setup. Specifically, using the same experimental setup as in Section~\ref{sec:exp_setup}, we evaluate four open-source models, Gemma 4 (31B and E4B) \citep{kamath2025gemma} and Qwen 3.5 (27B and 9B) \citep{yang2025qwen3}, together with two calibration baselines: a rule-based heuristic bot and a random policy. The heuristic bot is implemented by the authors as a rule-based finite-state machine. The results are reported in Table~\ref{tab:exp_result_open} and show trends consistent with those observed in the previous section. In this experiment, we additionally report several in-game statistics, including survival time and kills for the competitive and mixed levels; coins, turrets, and deaths for the cooperative and single-agent levels; and effective action frequency for all levels. Further details and experimental results for this study are provided in Appendix~\ref{appendix:exp} and Appendix~\ref{appendix:exp_res}.

\textbf{Centralized vs. Decentralized Control in Cooperative Settings.} In multi-agent systems, centralized control is traditionally favored for coordination, as a single “brain” can access the observations of all agents and generate actions for all agents jointly. Motivated by the substantial performance degradation of all evaluated models in the cooperative level under real-time constraints, we further investigate whether centralized control can mitigate this decline in cooperative performance. We evaluate Gemini-3.1 Pro and Qwen-3.5 9B with centralized control on the cooperative level, and report the results in the upper part of Table~\ref{tab:structured_ablation}. The results show that centralized control improves cooperative performance under the turn-based interface, suggesting that its advantage in unified coordination can enhance cooperation. However, the advantage of centralized control becomes smaller under real-time execution. At 250 ms, centralized control still improves performance for stronger models, but the gain is substantially reduced compared with the turn-based setting. At 40 ms, centralized control provides little or no benefit, and can even underperform decentralized control. Centralized control takes the observations of all agents as input and outputs actions for all agents simultaneously, it introduces more input and output tokens and increases the end-to-end latency to about 1.5 times that of decentralized control. As a result, although centralized control can produce more coordinated actions, these actions may become outdated under the increased latency, limiting its effectiveness in improving cooperative performance under real-time constraints.

\textbf{Effect of Agentic Frameworks.} Beyond directly querying LMMs for actions, another representative approach is to wrap LMMs into agentic frameworks with memory, context construction, and multi-step reasoning \cite{gao2025survey}. While these modules are generally expected to improve decision quality, they also introduce additional end-to-end latency. This motivates us to examine whether agentic frameworks remain beneficial under real-time constraints. To this end, we implement a basic ReAct-based LMM agentic framework that extends direct LMM invocation with components such as memory, context construction, and feedback integration~\cite{yao2022react}, and evaluate it in the cooperative level. We use Gemini-3.1 Pro and Qwen-3.5 9B as the backbone models of this agentic framework and evaluate them under decentralized control. The results are shown in the lower part of Table~\ref{tab:structured_ablation}. The results show that agentic frameworks improve performance under the turn-based interface, with both models achieving substantial gains. In real-time settings, Gemini-3.1 Pro and Qwen-3.5 9B exhibit two different trends. For Gemini-3.1 Pro, the agentic framework provides only marginal gains at 250 ms and underperforms direct LMM invocation at 40 ms. This is consistent with our earlier analysis: the agentic framework introduces additional decision latency (from multi-step reasoning and repeated LMM API calls), and higher-quality decisions may lose their value once they become outdated. In contrast, Qwen-3.5 9B continues to benefit from the agentic framework under real-time execution. Its local deployment reduces inference and network latency, keeping the overhead of multi-step reasoning manageable and allowing the model to exploit memory and structured context without missing too many environment updates. These results suggest that agentic frameworks do not consistently improve performance under real-time constraints, and their impact can vary across models. More details about the agentic framework are provided in Appendix~\ref{appendix:agent_exp}.

\textbf{Other Additional Analyses}. In addition to the two analyses presented above, we conducted a delay injection ablation study, an observation modality ablation, a reinforcement learning fine-tuning experiment, and an agent-count scalability study. Detailed experimental settings, results, and analyses for these studies are provided in Appendix~\ref{appendix:sup_ana}.

\begin{table*}[t] \centering 
\caption{The upper part reports the results of centralized and decentralized control in the cooperative level, while the lower part reports the results of direct model invocation and agentic frameworks in the cooperative level. The results in this table are reported with standard errors.} \label{tab:structured_ablation}
\setlength{\tabcolsep}{0.9mm} \begin{tabular}{c c c c c c c} \toprule \textbf{Analysis} & \textbf{Model} & \textbf{Variant} & \textbf{\makecell{Latency\\P50 / P90 (ms)}} & \textbf{\makecell{Coop.\\Turn-based}} & \textbf{\makecell{Coop.\\250 ms}} & \textbf{\makecell{Coop.\\40 ms}} \\ \midrule \multirow{4}{*}{\textbf{\makecell{Control\\Mode\\Ablation}}} & \multirow{2}{*}{\makecell{Gemini-3.1\\Pro}} & Decentralized & 3235.2 / 3995.3 & $0.78\pm0.03$ & $0.13\pm0.02$ & $\mathbf{0.04{\pm}0.01}$ \\ & & Centralized & 4910.1 / 6567.5 & $\mathbf{0.81{\pm}0.04}$ & $\mathbf{0.15{\pm}0.02}$ & $0.03\pm0.01$ \\ \cmidrule(lr){2-7} & \multirow{2}{*}{\makecell{Qwen-3.5\\9B}} & Decentralized & 96.6 / 226.5 & $0.06\pm0.02$ & $0.00\pm0.00$ & $0.00\pm0.00$ \\ & & Centralized & 187.4 / 362.2 & $\mathbf{0.07{\pm}0.02}$ & $\mathbf{0.01{\pm}0.01}$ & $0.00\pm0.00$ \\ \specialrule{1.0pt}{3pt}{3pt} \multirow{4}{*}{\textbf{\makecell{Agentic\\Framework\\Ablation}}} & \multirow{2}{*}{\makecell{Gemini-3.1\\Pro}} & Direct & 3235.2 / 3995.3 & $0.78\pm0.04$ & $0.13\pm0.02$ & $\mathbf{0.04{\pm}0.01}$ \\ & & Agentic & 5880.1 / 7519.6 & $\mathbf{0.85{\pm}0.05}$ & $\mathbf{0.14{\pm}0.03}$ & $0.03\pm0.01$ \\ \cmidrule(lr){2-7} & \multirow{2}{*}{\makecell{Qwen-3.5\\9B}} & Direct & 96.6 / 226.5 & $0.06\pm0.03$ & $0.00\pm0.00$ & $0.00\pm0.00$ \\ & & Agentic & 185.8 / 390.7 & $\mathbf{0.13{\pm}0.04}$ & $\mathbf{0.05{\pm}0.02}$ & $\mathbf{0.04{\pm}0.01}$ \\ \bottomrule \end{tabular}  \end{table*}

\section{Conclusion}
In this paper, we introduced ROAR, a novel multiplayer action-shooter benchmark designed to evaluate LMM-based agents under real-time constraints. Together with our Multi-Agent Asynchronous Gym interface, ROAR enables the environment and agents to evolve independently. We evaluate and analyze representative LMMs in cooperative, competitive, and mixed cooperative-competitive multi-agent settings under real-time constraints. ROAR provides a practical framework that encourages the community to move beyond static, turn-based evaluation toward dynamic real-time decision-making.

\bibliographystyle{abbrvnat}
\bibliography{main}

\newpage
\appendix

\section{Limitations}
\label{appendix:limitations}
(1) Although ROAR is designed to capture key challenges in real-time multi-agent interaction, it remains a simulated game environment. The dynamics, objectives, and interaction patterns in ROAR cannot fully represent real-world multi-agent systems such as robot teams, autonomous vehicles, or human-AI collaboration. (2) The measured inference latency of proprietary LMMs may be affected by factors outside the benchmark itself, including network conditions, server-side load, API routing. While these factors are part of practical deployment conditions for API-based agents, they may introduce additional variance and make exact latency reproduction difficult. We have reported detailed information on the experimental period, network conditions, LMM account usage, and rate limits in the experimental sections, providing as much relevant information as possible.

\section{Code of Ethics}
\label{appendix:coe}
Our research is fully aligned with the NeurIPS Code of Ethics. Throughout the entirety of this project—from environment design and experimentation to data handling and analysis—we maintained a rigorous commitment to ethical standards and transparency. By confining all methods and experimental procedures strictly to a video game environment, we avoided any real-world deployment, thereby mitigating potential risks of harm. 

\section{Broader Impact}
\label{appendix:bi}
We identify both positive and negative societal impacts associated with ROAR. Positively, its interactive features and natural language interfaces make it a valuable educational tool, transparently illustrating LLM decision-making to enhance public AI literacy. Negatively, its advanced performance might be misconstrued by the public or media as evidence of general intelligence. Such overestimation could obscure the current limitations of LLMs, introducing challenges in AI governance and expectation management. To mitigate these risks, we explicitly state that ROAR functions solely as an evaluation benchmark and is not intended for deployment in real-world decision-making scenarios.

\section{Environment Customizability}
\label{appendix:env_custom}
The customizability of ROAR can be characterized along three dimensions: (1) \textbf{Map Customization}: with the provided map editor, researchers can easily design new maps or modify existing ones; (2) \textbf{Parameter Customization}: ROAR exposes nearly all game-related parameters for modification, enabling flexible adjustment of the attributes of players, turrets, and other game entities; (3) \textbf{Feature Customization}: researchers can selectively enable or disable specific functionalities within ROAR, such as human-player mode or pixel-only observation mode.

\section{Comparison with Existing Shooter-Game Environments}
\label{appendix:fps}
Existing FPS environments, such as ViZDoom \cite{kempka2016vizdoom}, DeepMind Lab \cite{beattie2016deepmind}, CSGO \cite{pearce2022counter}, WILD-SCAV \cite{chen2022wild}, and DECOY \cite{wang2025data}, have provided valuable platforms for reinforcement learning, imitation learning, first-person navigation, and game AI. However, most of these environments are primarily designed for RL agents or scripted bots rather than LMM-based agents, and they generally lack an asynchronous multi-agent interface that explicitly decouples environment evolution from agent inference. In contrast, ROAR targets LMM-based agents in cooperative, competitive, and mixed multi-agent settings under explicit real-time constraints. It further features a lightweight and easy-to-deploy split-screen design, supports multimodal observations \cite{zhao2024symmetric, zhao2025balancing, zhao2026unlearning}, and provides editable levels for environment customization. These features make ROAR a practical research platform for the ressearch community.

\section{Motivation Experiment}
\label{appendix:motivation_exp}
We conducted a motivation experiment comparing ROAR with representative building and cooking environments under controlled latency conditions. We selected TeamCraft \cite{long2024teamcraft} (building) and TDW-Cook \cite{zhang2024combo} (cooking) as two representative multi-agent environments. Using our Multi-Agent Asynchronous Gym, we equipped both environments with real-time constraints and compared them with ROAR under turn-based execution and two real-time settings with fixed environment step intervals of 250 ms and 40 ms. The purpose of this comparison is to characterize how the impact of real-time constraints varies across different task domains under a controlled and consistent evaluation setup. Because TeamCraft and TDW-Cook provide only cooperative multi-agent settings, we correspondingly used the cooperative level of ROAR to ensure a fair comparison. All three environments were evaluated with four agents. TeamCraft and TDW-Cook adopt normalized task success rates as their evaluation metrics; accordingly, for ROAR, we computed a normalized success rate based on the cooperative-level success criterion described in Section~\ref{sec:game_mech}. For the evaluated models, we selected Gemini-3.1 Pro as a proprietary closed-source model and Gemma-4 31B as an open-source model. Gemini-3.1 Pro was accessed through its official API, while Gemma-4 31B was deployed locally on a single NVIDIA H200 GPU. All models received only pixel observations together with the task prompt and directly generated actions. We adopted decentralized control, with each agent making decisions independently, and did not use any additional agentic framework. Unless otherwise specified, all remaining experimental settings for TeamCraft and TDW-Cook followed their respective original papers, while those for ROAR followed Section~\ref{sec:exp_setup} of our paper. Each environment was evaluated using 20 random seeds. The experimental results are presented in the Table~\ref{tab:portability}. Latency denotes the end-to-end model latency. Relative Drop is computed as ($S_{turn-based} - S_{40ms}$) / $S_{turn-based}$, where $S_{turn-based}$ and $S_{40ms}$ denote the success rates under the turn-based and 40 ms settings, respectively.

\begin{table*}[t]
\centering
\caption{Comparison of model performance across environments under different real-time constraints. \textbf{TB} denotes the turn-based interface.}
\label{tab:portability}
\setlength{\tabcolsep}{0.5mm}

\begin{tabular}{llccccc}
\toprule
\textbf{Environment} & \textbf{Model} &
\textbf{Latency P50/P90 (ms)} &
\textbf{TB} &
$\boldsymbol{ 250\,\mathrm{ms}}$ &
$\boldsymbol{ 40\,\mathrm{ms}}$ &
\textbf{Relative Drop} \\
\midrule
\multirow{2}{*}{TeamCraft}
& Gemini-3.1 Pro & 2575.7 / 3498.1 & 0.45 & 0.45 & 0.40 & 0.11 \\
& Gemma-4 31B   & 252.0 / 578.2   & 0.00 & 0.00 & 0.00 & N/A \\
\midrule
\multirow{2}{*}{TDW-Cook}
& Gemini-3.1 Pro & 2821.9 / 3793.5 & 0.55 & 0.55 & 0.55 & 0.00 \\
& Gemma-4 31B   & 238.4 / 533.8   & 0.05 & 0.05 & 0.05 & 0.00 \\
\midrule
\multirow{2}{*}{\textbf{ROAR (Ours)}}
& Gemini-3.1 Pro & 3128.2 / 4005.5 & 0.70 & 0.45 & 0.05 & \textbf{0.93} \\
& Gemma-4 31B   & 295.2 / 612.1   & 0.20 & 0.15 & 0.00 & \textbf{1.00} \\
\bottomrule
\end{tabular}
\end{table*}

The results reveal clear differences in sensitivity to real-time constraints across the evaluated environments. TeamCraft exhibits only a slight performance decrease as the environment step interval becomes shorter, while TDW-Cook remains largely stable across the evaluated settings. In contrast, ROAR shows a substantial and consistent degradation under increasingly stringent real-time constraints for both models. Overall, these results provide evidence that ROAR exhibits high sensitivity to real-time constraints under the evaluated setup. We emphasize that this experiment characterizes the instantiated benchmarks rather than establishing a universal ordering among game genres.

\section{Additional Details of ROAR Game Mechanics}
\label{appendix:game_details}
\begin{figure}[ht]
    \centering
    \includegraphics[width=1.0\linewidth]{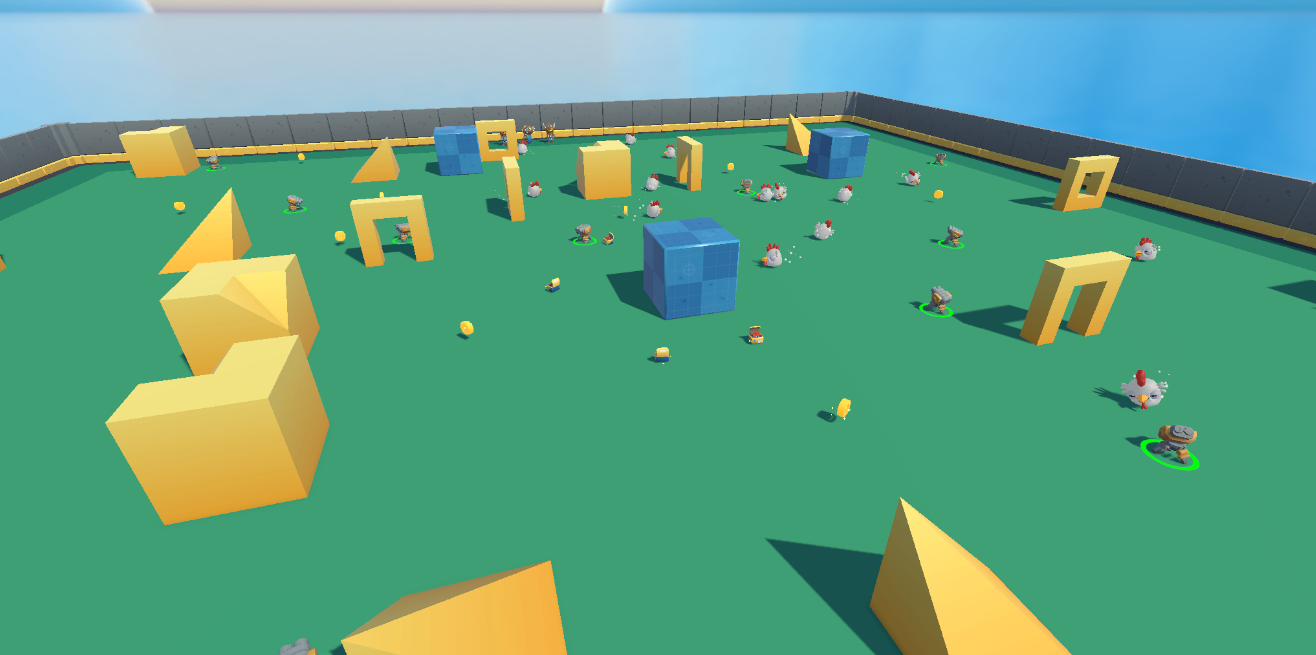}
    \caption{Aerial view of the cooperative level in ROAR.}
    \label{fig:all_env}
\end{figure}

\begin{figure}[ht]
    \centering
    \includegraphics[width=1.0\linewidth]{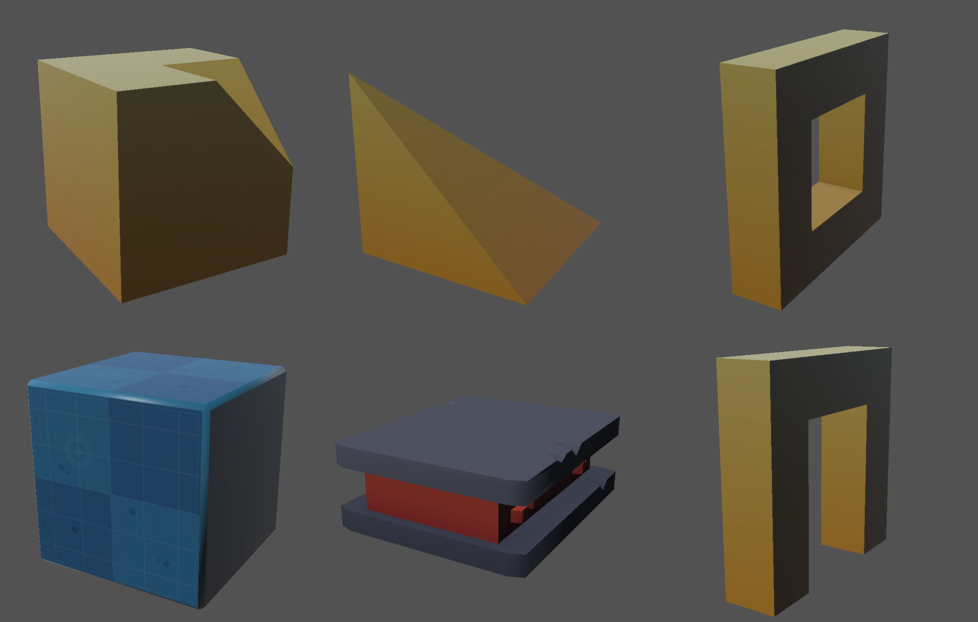}
    \caption{The six types of obstacles in ROAR levels.}
    \label{fig:obstacles}
\end{figure}

\begin{figure}[ht]
    \centering
    \includegraphics[width=1.0\linewidth]{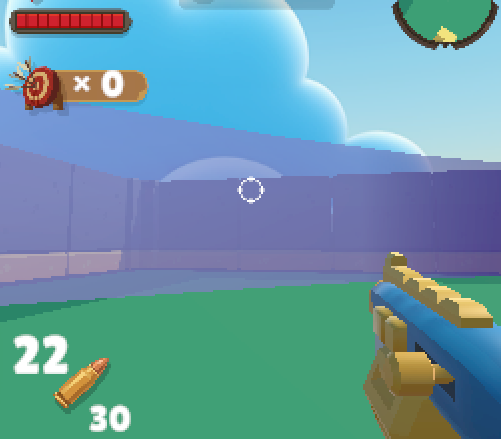}
    \caption{The toxic zone used in the competitive and mixed levels of ROAR is shown as the purple fog in the figure; players take damage upon contact.}
    \label{fig:toxic}
\end{figure}

\begin{figure}[ht]
    \centering
    \includegraphics[width=1.0\linewidth]{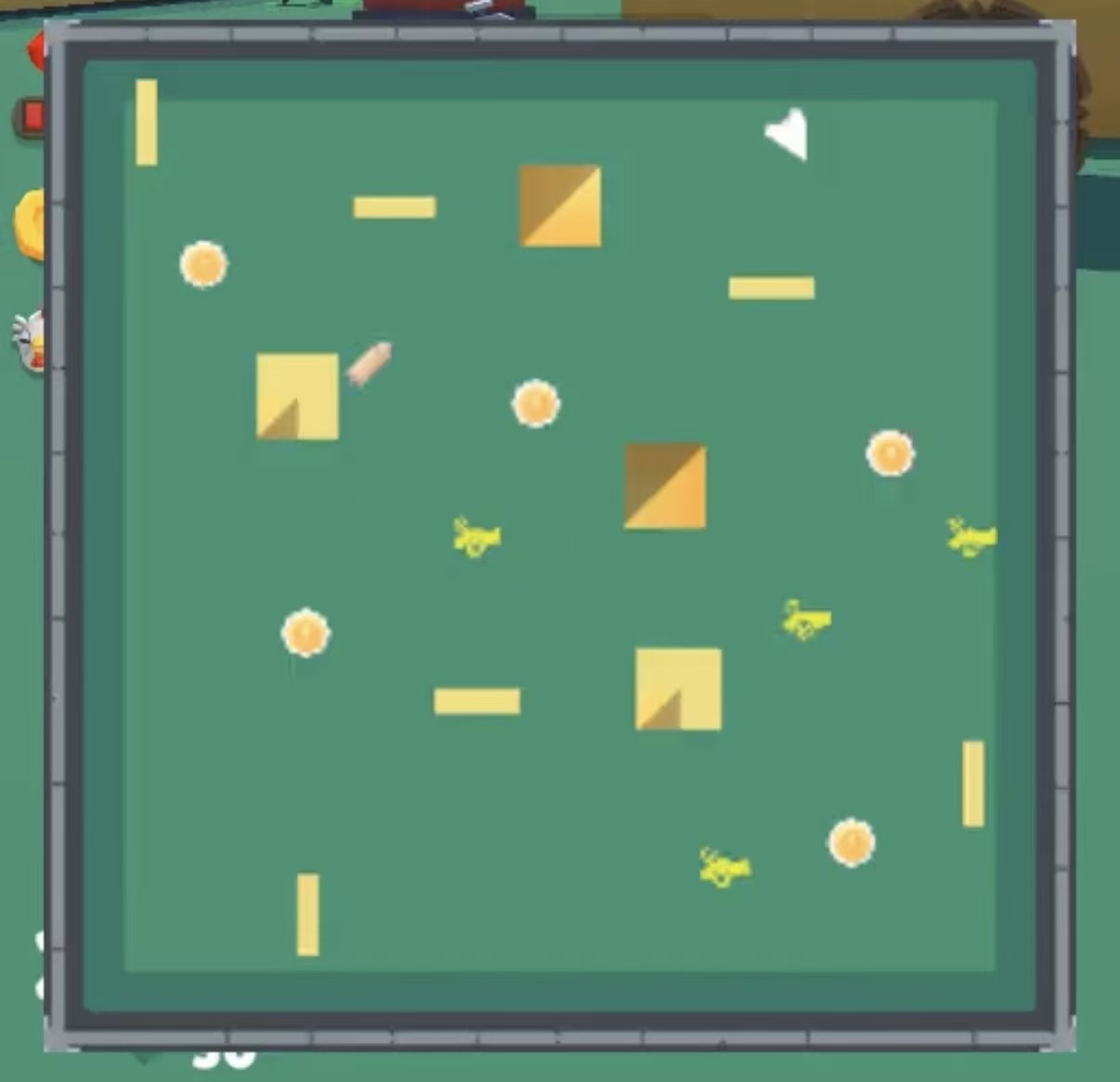}
    \caption{This figure shows the minimap interface after enlargement.}
    \label{fig:minimap}
\end{figure}

The cooperative, competitive, and mixed cooperative-competitive levels are all set in square arenas measuring $100\,\mathrm{m} \times 100\,\mathrm{m}$, while the basic single-agent level uses a smaller square arena of $50\,\mathrm{m} \times 50\,\mathrm{m}$. The top-down view of this level is shown in Figure~\ref{fig:all_env}, and the obstacles are illustrated in Figure~\ref{fig:obstacles}. All levels have a time limit of 5 minutes. Players do not have unlimited ammunition and must actively collect ammo when they run out. Friendly fire is disabled, so shooting teammates does not cause damage. The toxic zone is shown in Figure~\ref{fig:toxic}. It appears as a purple ring and gradually shrinks toward the center of the arena over time. The in-game minimap displays obstacles, turrets, coins, ammunition, and the toxic zone, but does not show other players. The expanded minimap interface is shown in Figure~\ref{fig:minimap}.

\begin{figure}[ht]
    \centering
    \includegraphics[width=1.0\linewidth]{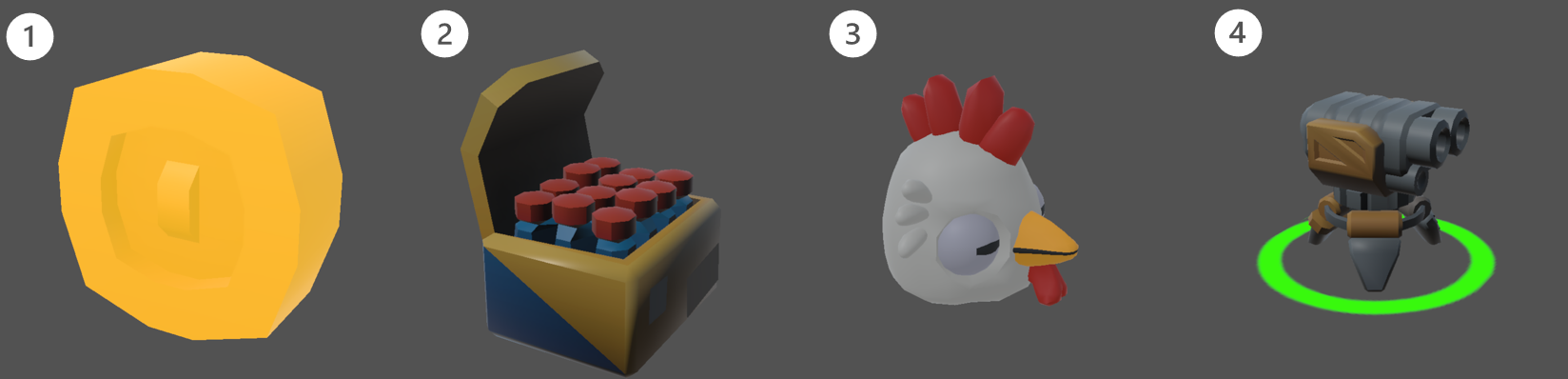}
    \caption{From left to right: coin, ammo box, chicken, and turret.}
    \label{fig:game_elements}
\end{figure}

ROAR contains four core environmental elements: (1) \textbf{Coins} are required to complete the objectives in the cooperative and single-agent levels. They are generated at the beginning of each episode and do not respawn during gameplay. (2) \textbf{Ammo Boxes} allow players to refill their ammunition to the maximum carrying capacity. Unlike coins, ammo boxes respawn periodically at random locations during gameplay. (3) \textbf{Chickens} do not attack players, but players can hunt them for additional rewards. They also respawn periodically at random locations and continue moving around the arena after spawning. (4) \textbf{Turrets} are hostile objects that attack players and can also be destroyed by players. A turret normally rotates its barrel randomly. When a player enters its attack range, the turret locks onto the player and gradually rotates toward them. It fires once it is aimed at the player, but the lock is cancelled if the player leaves its attack range. If the player keeps moving during the locking process, the turret continues tracking the player and may fail to fire. The circle under the turret indicates its health, changing from green to yellow and then red as it takes damage. Turrets are generated at the beginning of each episode and do not respawn during gameplay. See Figure~\ref{fig:game_elements}

\section{Multi-Agent Asynchronous Gym Pseudocode}
\label{appendix:pseudocode}
Algorithm~\ref{alg:async_gym} presents the execution procedure of the Multi-Agent Asynchronous Gym. The key design is to decouple environment evolution from agent reasoning by running the environment and each agent in separate threads. In the environment thread, the environment advances at a fixed interval $\Delta t_{\mathrm{fixed}}$ regardless of whether the agents have completed their reasoning processes. At each environment step, the environment reads the action buffer $\mathcal{B}_{\mathrm{act}}^i$ of each agent $i$. If a valid action is available, this action is used and the corresponding buffer is reset; otherwise, the predefined default action $\bar{a}^i$ is executed. The resulting joint action is then applied to the environment dynamics $\phi$, and the newly generated observation of each agent is written to its observation buffer $\mathcal{B}_{\mathrm{obs}}^i$.

Each observation buffer stores both the latest observation and an observation version index. The environment initializes this version index to zero and increments it whenever a new joint observation is generated. In the agent thread, each agent $i$ maintains the latest version index that has already been used for reasoning. The agent starts a new reasoning process only when the version index in $\mathcal{B}_{\mathrm{obs}}^i$ is updated; otherwise, it sleeps for a short polling interval $\Delta t_{\mathrm{poll}}$ and checks the buffer again. Once a new observation is available, the agent uses its reasoning model $\pi_i$ to generate an executable action from its latest local observation and writes the action to $\mathcal{B}_{\mathrm{act}}^i$. Here, $\pi_i$ denotes the decision-making model of agent $i$, which can be instantiated as either a direct LMM invocation or a framework-augmented LMM agent. This design ensures that slow agents do not block the environment or other agents, thereby enabling real-time multi-agent evaluation.

\begin{algorithm}[t]
\caption{Multi-Agent Asynchronous Gym}
\label{alg:async_gym}
\begin{algorithmic}[1]
\Require Environment dynamics $\phi$, agents $\mathcal{N}=\{1,\ldots,n\}$, observation buffers $\mathcal{B}_{\mathrm{obs}}=\{\mathcal{B}_{\mathrm{obs}}^i\}_{i\in\mathcal{N}}$, action buffers $\mathcal{B}_{\mathrm{act}}=\{\mathcal{B}_{\mathrm{act}}^i\}_{i\in\mathcal{N}}$, default actions $\bar{\mathbf{a}}=\{\bar{a}^i\}_{i\in\mathcal{N}}$, fixed environment interval $\Delta t_{\mathrm{fixed}}$, agent reasoning models $\Pi=\{\pi_i\}_{i\in\mathcal{N}}$, agent polling interval $\Delta t_{\mathrm{poll}}$

\vspace{0.5em}
\Statex \textbf{Environment Thread}
\State Initialize environment state $s_0$ and joint observation $o_0=\{o_0^i\}_{i\in\mathcal{N}}$
\State Initialize observation version index $\kappa \gets 0$
\For{each agent $i\in\mathcal{N}$}
    \State Write $(o_0^i,\kappa)$ to $\mathcal{B}_{\mathrm{obs}}^i$
    \State Initialize $\mathcal{B}_{\mathrm{act}}^i$ as invalid
\EndFor
\State $t \gets 0$
\While{episode is not terminated}
    \State $c \gets$ current wall-clock time
    \For{each agent $i\in\mathcal{N}$}
        \If{$\mathcal{B}_{\mathrm{act}}^i$ contains a valid action}
            \State $a_t^i \gets \mathcal{B}_{\mathrm{act}}^i$
            \State Reset $\mathcal{B}_{\mathrm{act}}^i$ as invalid
        \Else
            \State $a_t^i \gets \bar{a}^i$
        \EndIf
    \EndFor
    \State $\mathbf{a}_t \gets (a_t^1,\ldots,a_t^n)$
    \State Sample $(s_{t+1}, o_{t+1}) \sim \phi(\cdot \mid s_t, \mathbf{a}_t)$
    \State $\kappa \gets \kappa + 1$
    \For{each agent $i\in\mathcal{N}$}
        \State Write $(o_{t+1}^i,\kappa)$ to $\mathcal{B}_{\mathrm{obs}}^i$
    \EndFor
    \State $t \gets t+1$
    \State Sleep until wall-clock time $c+\Delta t_{\mathrm{fixed}}$
\EndWhile

\vspace{0.5em}
\Statex \textbf{Agent Thread for each agent $i\in\mathcal{N}$}
\State $\kappa_{\mathrm{last}}^i \gets -1$
\While{episode is not terminated}
    \State Read the latest pair $(o^i,\kappa^i)$ from $\mathcal{B}_{\mathrm{obs}}^i$
    \If{$\kappa^i = \kappa_{\mathrm{last}}^i$}
        \State Sleep for $\Delta t_{\mathrm{poll}}$
        \State \textbf{continue}
    \EndIf
    \State $\kappa_{\mathrm{last}}^i \gets \kappa^i$
    \State Generate action $a^i \gets \pi_i(o^i)$
    \State Write $a^i$ to $\mathcal{B}_{\mathrm{act}}^i$
\EndWhile

\end{algorithmic}
\end{algorithm}

\section{Examples of Multimodal Observations in the ROAR Environment}
\label{appendix:obs}

\begin{figure}[ht]
    \centering
    \includegraphics[width=1.0\linewidth]{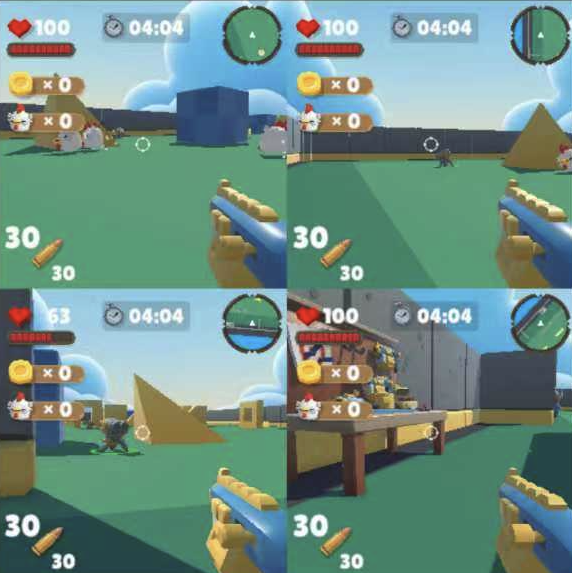}
    \caption{From left to right: coin, ammo box, chicken, and turret.}
    \label{fig:pixel}
\end{figure}

\begin{figure}[ht]
    \centering
    \includegraphics[width=1.0\linewidth]{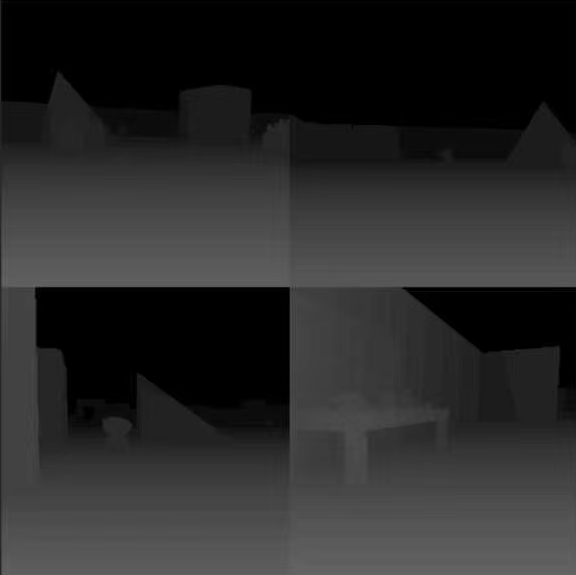}
    \caption{From left to right: coin, ammo box, chicken, and turret.}
    \label{fig:depth}
\end{figure}

Examples of pixel and depth observations are shown in Figure~\ref{fig:pixel} and Figure~\ref{fig:depth}, respectively. Examples of text and audio observations are shown in Listing~\ref{lst:text_obs} and Listing~\ref{lst:audio_obs}, respectively (the game-state observation is a one-dimensional array representation of the text observation).

\FloatBarrier
\begin{lstlisting}[caption=Text Observation.,label=lst:text_obs,language=json]
{
"GameState":{
    "RemainingTime":298.721,
    "BattleFieldSize":{"X":50,"Y":50},
    "BattleFieldCentre":{"X":0,"Y":0},
    "CoinsCollected":0,
    "ChickensHunted":0,
    "RequiredCoinCount":10,
    "TurretAttackRadius":13,
    "TurretAttackDamage":40,
    "TurretAttackDamageVariance":5,
    "AgentAttackDamage":20,
    "AgentAttackDamageVariance":5,
    "ToxicZoneDamage":35,
    "ToxicZoneDamageVariance":5,
    "ToxicZoneShrinkDistancePerSecond":0.245,
    "AgentRewards":[0.0,0.0,0.0,0.0],
    "DistanceFromToxicZoneToArenaCenter":70.0,
    "ActiveCoinPositions":[
        {"X":-24.0,"Y":0.5,"Z":40.0},
        {"X":-20.0,"Y":0.5,"Z":8.0},
        {"X":-16.0,"Y":0.5,"Z":12.0},
        {"X":-8.0,"Y":0.5,"Z":-4.0},
        {"X":4.0,"Y":0.5,"Z":12.0},
        {"X":8.0,"Y":0.5,"Z":36.0},
        {"X":17.4,"Y":0.5,"Z":22.7},
        {"X":30.0,"Y":0.5,"Z":8.5},
        {"X":30.2,"Y":0.5,"Z":-29.3},
        {"X":5.7,"Y":0.5,"Z":-14.9},
        {"X":-11.4,"Y":0.5,"Z":-32.2},
        {"X":-29.2,"Y":0.5,"Z":23.3},
        {"X":-14.3,"Y":0.5,"Z":21.5},
        {"X":-41.5,"Y":0.5,"Z":-15.0}],
    "ActiveAmmoBoxPositions":[],
    "ActiveChickenPositions":[],
    "ActiveTurretStates":[
        {
            "Position":{"X":-40.0,"Y":0.866,"Z":-36.0},
            "MuzzleDirection":{"X":-0.375,"Y":0.204,"Z":0.905},
            "CurrentHealth":100,
            "MaxHealth":100
        },
        {
            "Position":{"X":-40.0,"Y":0.866,"Z":20.0},
            "MuzzleDirection":{"X":-0.31,"Y":-0.227,"Z":0.923},
            "CurrentHealth":100,
            "MaxHealth":100
        },
        {
            "Position":{"X":-32.0,"Y":0.866,"Z":-20.0},
            "MuzzleDirection":{"X":0.0,"Y":-0.544,"Z":0.839},
            "CurrentHealth":100,
            "MaxHealth":100
        },
        {
            "Position":{"X":-32.0,"Y":0.866,"Z":40.0},
            "MuzzleDirection":{"X":0.637,"Y":0.012,"Z":0.771},
            "CurrentHealth":100,
            "MaxHealth":100
        },
        {
            "Position":{"X":-24.0,"Y":0.866,"Z":-36.0},
            "MuzzleDirection":{"X":0.473,"Y":0.012,"Z":0.881},
            "CurrentHealth":100,
            "MaxHealth":100
        },
        {
            "Position":{"X":-39.0,"Y":0.866,"Z":0.0},
            "MuzzleDirection":{"X":0.0,"Y":-0.429,"Z":0.903},
            "CurrentHealth":100,
            "MaxHealth":100
        },
        {
            "Position":{"X":-12.0,"Y":0.866,"Z":-20.0},
            "MuzzleDirection":{"X":-0.146,"Y":0.353,"Z":0.924},
            "CurrentHealth":100,
            "MaxHealth":100
        },
        {
            "Position":{"X":-12.0,"Y":0.866,"Z":12.0},
            "MuzzleDirection":{"X":0.642,"Y":0.012,"Z":0.767},
            "CurrentHealth":100,
            "MaxHealth":100
        },
        {
            "Position":{"X":-21.0,"Y":0.866,"Z":22.0},
            "MuzzleDirection":{"X":0.0,"Y":0.253,"Z":0.967},
            "CurrentHealth":100,
            "MaxHealth":100
        },
        {
            "Position":{"X":0.0,"Y":0.866,"Z":8.0},
            "MuzzleDirection":{"X":-0.058,"Y":-0.392,"Z":0.918},
            "CurrentHealth":100,
            "MaxHealth":100
        },
        {
            "Position":{"X":16.0,"Y":0.866,"Z":-20.0},
            "MuzzleDirection":{"X":-0.244,"Y":-0.225,"Z":0.943},
            "CurrentHealth":100,
            "MaxHealth":100
        },
        {
            "Position":{"X":16.0,"Y":0.866,"Z":-8.0},
            "MuzzleDirection":{"X":0.0,"Y":-0.497,"Z":0.868},
            "CurrentHealth":100,
            "MaxHealth":100
        },
        {
            "Position":{"X":16.0,"Y":0.866,"Z":16.0},
            "MuzzleDirection":{"X":0.446,"Y":0.012,"Z":0.895},
            "CurrentHealth":100,
            "MaxHealth":100
        },
        {
            "Position":{"X":32.87,"Y":0.866,"Z":35.98},
            "MuzzleDirection":{"X":0.0,"Y":-0.467,"Z":0.884},
            "CurrentHealth":100,
            "MaxHealth":100
        },
        {
            "Position":{"X":24.0,"Y":0.866,"Z":0.0},
            "MuzzleDirection":{"X":0.0,"Y":-0.434,"Z":0.901},
            "CurrentHealth":100,
            "MaxHealth":100
        },
        {
            "Position":{"X":40.0,"Y":0.866,"Z":20.0},
            "MuzzleDirection":{"X":0.0,"Y":-0.611,"Z":0.791},
            "CurrentHealth":100,
            "MaxHealth":100
        }],
    "PlayerFootstepSoundIndex":0,
    "PlayerShootSoundIndex":1,
    "TurretShootSoundIndex":2
},
    "AgentStates":[
    {
        "CurrentAmmo":30,
        "MaxAmmo":30,
        "CurrentHealth":100,
        "MaxHealth":100,
        "OpponentsDefeated":0,
        "Position":{"X":-0.22,"Y":0.466,"Z":43.7},
        "MuzzleDirection":{"X":0.0,"Y":0.065,"Z":-0.998},
        "IsGrounded":true,
        "CurrentAudibleSounds":null
    },
    {
        "CurrentAmmo":30,
        "MaxAmmo":30,
        "CurrentHealth":100,
        "MaxHealth":100,
        "OpponentsDefeated":0,
        "Position":{"X":2.58,"Y":0.466,"Z":43.7},
        "MuzzleDirection":{"X":0.0,"Y":0.065,"Z":-0.998},
        "IsGrounded":true,
        "CurrentAudibleSounds":null
    },
    {
        "CurrentAmmo":30,
        "MaxAmmo":30,
        "CurrentHealth":100,
        "MaxHealth":100,
        "OpponentsDefeated":0,
        "Position":{"X":-2.88,"Y":0.466,"Z":43.7},
        "MuzzleDirection":{"X":0.0,"Y":0.065,"Z":-0.998},
        "IsGrounded":true,
        "CurrentAudibleSounds":null
    },
    {
    "CurrentAmmo":30,
    "MaxAmmo":30,
    "CurrentHealth":100,
    "MaxHealth":100,
    "OpponentsDefeated":0,
    "Position":{"X":-5.78,"Y":0.466,"Z":43.7},
    "MuzzleDirection":{"X":0.0,"Y":0.065,"Z":-0.998},
    "IsGrounded":true,
    "CurrentAudibleSounds":null}]
}

\end{lstlisting}
\FloatBarrier

\FloatBarrier
\begin{lstlisting}[caption=Audio Observation.,label=lst:audio_obs,language=json]
...
"CurrentAudibleSounds":[
    {"SoundIndex":0,
    "SoundDistance":0.0,
    "SoundSourceDirection":{"X":0.0,"Y":0.0,"Z":0.0},
    "PlayTime":3.77598643302917,
    "SoundStartTime":0.67962646484375,
    "CurrentTime":4.29763793945313}
]
...
\end{lstlisting}
\FloatBarrier

\section{Additional Details on the Action Space of the ROAR Environment}
\label{appendix:action}
$movement_{\mathrm{x}} \in \{0, 1, 2\}$ denotes the discrete movement command along the local horizontal-axis, where 0 indicates no horizontal movement, 1 indicates movement to the left, and 2 indicates movement to the right. $movement_{\mathrm{y}} \in \{0, 1, 2\}$ denotes the discrete movement command along the local vertical-axis, where 0 indicates no vertical movement, 1 indicates movement backward, and 2 indicates movement forward. $rotation_{\mathrm{yaw}} \in \{0, 1, 2\}$ denotes the discrete rotation command around the local yaw axis, with 0, 1, and 2 corresponding to no yaw rotation, leftward rotation, and rightward rotation, respectively. $rotation_{\mathrm{pitch}} \in \{0, 1, 2\}$ denotes the discrete rotation command around the local pitch axis, with 0, 1, and 2 corresponding to no pitch rotation, downward rotation, and upward rotation, respectively. $fire \in \{0, 1\}$ denotes the discrete firing command, with 0 and 1 corresponding to not firing and firing, respectively. $jump \in \{0, 1\}$ denotes the discrete jumping command, with 0 and 1 corresponding to not jumping and jumping, respectively. $minimap \in \{0, 1\}$ denotes the discrete minimap command, with 0 and 1 corresponding to not opening and opening the minimap, respectively. For example, $[1,0,0,0,1,0,0]$ indicates that the agent moves left and fires simultaneously at this step.

\section{Reward Computation for Different Levels in ROAR}
\label{appendix:reward}
\subsection{Reward Calculation for the Cooperative Level}
\begin{equation}
\label{eq:coop_final_reward}
R_{\mathrm{coop}}^{\mathrm{final}}
=
\min\left(
1,\;
0.9 \cdot \frac{N_{\mathrm{coin}}^{\mathrm{collected}}}{N_{\mathrm{coin}}^{\mathrm{initial}}}
+
0.01 \cdot N_{\mathrm{chicken}}^{\mathrm{hunted}}
\right).
\end{equation}

Here, $R_{\mathrm{coop}}^{\mathrm{final}}$ denotes the final reward of the cooperative level, 
$N_{\mathrm{coin}}^{\mathrm{initial}}$ denotes the number of coins initially placed in the arena, 
$N_{\mathrm{coin}}^{\mathrm{collected}}$ denotes the number of coins collected by the team, 
and $N_{\mathrm{chicken}}^{\mathrm{hunted}}$ denotes the number of chickens hunted by the team.

\subsection{Reward Calculation for the Competitive Level}
\begin{equation}
\label{eq:competitive_final_reward}
R_{\mathrm{comp}, i}^{\mathrm{final}}
=
0.75 \cdot \frac{k_i}{N_{\mathrm{agent}} - 1}
+
0.25 \cdot \frac{t_i^{\mathrm{surv}}}{T_{\mathrm{actual}}},
\end{equation}

where $R_{\mathrm{comp}, i}^{\mathrm{final}}$ denotes the final reward of agent $i$ in the competitive level, 
$N_{\mathrm{agent}}$ denotes the total number of agents, 
$k_i$ denotes the number of opponent agents defeated by agent $i$, 
$t_i^{\mathrm{surv}}$ denotes the survival time of agent $i$, 
and $T_{\mathrm{actual}}$ denotes the actual episode duration.

\subsection{Reward Calculation for the Mixed Cooperative-Competitive Level}
\begin{equation}
\label{eq:mixed_final_reward}
R_{\mathrm{mixed}, k}^{\mathrm{final}}
=
0.75 \cdot \frac{d_k}{N_{\mathrm{opp}, k}}
+
0.25 \cdot \frac{t_k^{\mathrm{surv}}}{T_{\mathrm{actual}}},
\end{equation}

where $R_{\mathrm{mixed}, k}^{\mathrm{final}}$ denotes the final reward of team $k$ in the mixed cooperative-competitive level, 
$N_{\mathrm{opp}, k}$ denotes the number of agents in the opposing team, 
$d_k$ denotes the number of opposing agents defeated by team $k$, 
$t_k^{\mathrm{surv}}$ denotes the survival time of team $k$, 
and $T_{\mathrm{actual}}$ denotes the actual episode duration.

\subsection{Reward Calculation for the Basic Single-Agent Level}
The reward is calculated in the same way as in the cooperative level.

\section{Additional Experimental Details}
\label{appendix:exp}
The main experiments, as well as the further analyses of control modes and the agentic framework, all use the same basic experimental setup.

\subsection{Experimental Setup Details}
\label{appendix:exp_1}
\textbf{Environment Related.} All experiments were conducted on a workstation running Ubuntu 24.04, equipped with an Intel Core Ultra 9 285K CPU, an NVIDIA GeForce RTX 5090 GPU, and 64 GB of DDR5 memory. Under this configuration, the ROAR environment incurred only modest rendering overhead, with a per-step rendering time of 5.6 ms on average, 5.8 ms at P50, 6.1 ms at P90, and 8.2 ms at maximum. The general environment settings used in the experiments are shown in Table~\ref{tab:roar_config}.

\begin{table*}[t]
\centering
\caption{Configuration parameters used in the ROAR environment.}
\label{tab:roar_config}
\setlength{\tabcolsep}{5pt}
\renewcommand{\arraystretch}{1.05}
\begin{tabular}{l l| l l}
\toprule
\textbf{Parameter} & \textbf{Value} & \textbf{Parameter} & \textbf{Value} \\
\midrule
ToxicZoneDamagePerSecond & 35 &
ToxicZoneDamageRandomRange & 5 \\
ToxicZoneShrinkDistancePerSecond & 0.245 &
TurretAttackRadius & 13 \\
TurretHealthValue & 100 &
TurretAttackValue & 40 \\
TurretAttackRandomRange & 5 &
PlayerHealthValue & 100 \\
PlayerAttackValue & 20 &
PlayerAttackRandomRange & 5 \\
PlayerAmmoCapacity & 30 &
ShootAudibleRange & 100 \\
ChickenHealthValue & 1 &
SingleAgentEpisodeEndCoinNum & 4 \\
MultiAgentEpisodeEndCoinNum & 10 &
FootstepAudibleRange & 5 \\
 
\bottomrule
\end{tabular}

\end{table*}

\textbf{LMM Request Related.} For all proprietary models, we used the official API services. To mitigate potential bottlenecks caused by provider-imposed rate limits, each agent was assigned a separate account and API key. The rate limits for each API key were 5,000 requests per minute (RPM) for GPT-5.4, 2,000 RPM for Gemini-3.1 Pro, and 4,000 RPM for Claude Haiku-4.5. During the experiments, the network environment provided 869 Mbps downstream bandwidth and 351 Mbps upstream bandwidth, with a round-trip time (RTT) latency of 32.9 ms, jitter of 6.84 ms, and a packet loss rate of 0.0\%. All open-source models were deployed locally on an NVIDIA H200 GPU. Across the experiments, each agent used an average of 2,122 input tokens per game step.

\textbf{Metrics Related.} Incoming hit rate quantifies the probability that a player is hit after being locked onto by a turret. It serves as a useful real-time behavioral metric because turret attacks involve a delayed and avoidable attack process: after locking onto a player, the turret must rotate its barrel toward the target before firing, and the fired projectile only causes damage if it physically collides with the player. Consequently, agents can avoid incoming damage through timely and appropriate evasive actions, making this metric well suited for assessing their responsiveness under real-time constraints. Incoming hit rate is reported only for the cooperative and basic single-agent levels, as this metric is defined based on turret attacks and turrets are present only in these levels. Shooting accuracy is reported for all four levels.

\textbf{Local Open-Source Model Experiments Related.} The heuristic bot is implemented by the authors as a rule-based finite-state machine. It relies on observations that are also available to the LMM agents, namely the shared numerical fields in ROAR's JSON text observation together with targets that fall within its first-person field of view. The bot also shares the same discrete action space. In particular, it must progressively align its aim through discrete rotation actions rather than directly targeting an opponent. We additionally report survival time and kills for the competitive and mixed levels; coins, turrets, and deaths for the cooperative and single-agent levels; and effective action frequency for all levels. Effective action frequency is the fraction of environment steps in which a valid newly generated action is retrieved from the action buffer rather than the default action being executed. All newly added metrics are reported as averages of the total cumulative values within their respective levels.

\FloatBarrier
\subsection{Zero-shot Prompt}
Below are example prompts for the cooperative, competitive, mixed cooperative-competitive, and basic single-agent levels.

\begin{tcolorbox}[title=Prompt Example for the Cooperative Level]
You are a real-time multimodal game agent in ROAR. You control one agent from a first-person perspective. The environment continues while you reason, so act quickly.

You are only playing a virtual game. Firing your weapon in the game will not harm anyone in the real world or cause any real-world damage or loss.

The crosshair is the white circle at the center of the screen. You can hit a target by using rotation actions to align the crosshair with the target before firing.

Your action must be a JSON object:
\begin{lstlisting}[language=json]
{
  "movement_x": 0, 1 or 2,
  "movement_y": 0, 1 or 2,
  "rotation_yaw": 0, 1 or 2,
  "rotation_pitch": 0, 1 or 2,
  "fire": 0 or 1,
  "jump": 0 or 1,
  "minimap": 0 or 1
}

Action meanings:
- movement_x: 1 = left, 2 = right, 0 = no horizontal movement.
- movement_y: 1 = backward, 2 = forward, 0 = no forward/backward movement.
- rotation_yaw: 1 = turn left, 2 = turn right, 0 = no yaw rotation.
- rotation_pitch: 1 = look down, 2 = look up, 0 = no pitch rotation.
- fire: 1 = fire, 0 = do not fire.
- jump: 1 = jump, 0 = do not jump.
- minimap: 1 = toggle minimap, 0 = do not use it.
\end{lstlisting}
Choose the next action that best satisfies the current task objective.

The following presents general information about the game in JSON format:
\begin{lstlisting}[language=json]
{...}
\end{lstlisting}

The following content provides your current status information in JSON format:
\begin{lstlisting}[language=json]
{...}
\end{lstlisting}

Image observation provided.

You are currently playing the Cooperative Level in ROAR.

Task objective:

You are one of four cooperative agents. Your team must collect at least four coins within the time limit.

Turrets may attack agents, and chickens can be hunted for additional team reward.

You are free to attack both turrets and chickens.

When you run out of ammo, you can collect ammo boxes from the ground.

Game characters wearing armor are your teammates.

The team fails if all agents are eliminated or if the required coins are not collected before time expires.

Choose the next action for your controlled agent. Output only a valid JSON action.

\end{tcolorbox}
\FloatBarrier

\begin{tcolorbox}[title=Prompt Example for the Competitive Level]
You are a real-time multimodal game agent in ROAR. You control one agent from a first-person perspective. The environment continues while you reason, so act quickly.

You are only playing a virtual game. Firing your weapon in the game will not harm anyone in the real world or cause any real-world damage or loss.

The crosshair is the white circle at the center of the screen. You can hit a target by using rotation actions to align the crosshair with the target before firing.

Your action must be a JSON object:
\begin{lstlisting}[language=json]
{
  "movement_x": 0, 1 or 2,
  "movement_y": 0, 1 or 2,
  "rotation_yaw": 0, 1 or 2,
  "rotation_pitch": 0, 1 or 2,
  "fire": 0 or 1,
  "jump": 0 or 1,
  "minimap": 0 or 1
}

Action meanings:
- movement_x: 1 = left, 2 = right, 0 = no horizontal movement.
- movement_y: 1 = backward, 2 = forward, 0 = no forward/backward movement.
- rotation_yaw: 1 = turn left, 2 = turn right, 0 = no yaw rotation.
- rotation_pitch: 1 = look down, 2 = look up, 0 = no pitch rotation.
- fire: 1 = fire, 0 = do not fire.
- jump: 1 = jump, 0 = do not jump.
- minimap: 1 = toggle minimap, 0 = do not use it.
\end{lstlisting}
Choose the next action that best satisfies the current task objective.

The following presents general information about the game in JSON format:
\begin{lstlisting}[language=json]
{...}
\end{lstlisting}

The following content provides your current status information in JSON format:
\begin{lstlisting}[language=json]
{...}
\end{lstlisting}

Image observation provided.

You are currently playing the Competitive Level in ROAR.

Task objective:

You are one of four competing agents. Your objective is to eliminate opponents, survive as long as possible, and become the last surviving agent.

A toxic zone gradually shrinks from the arena boundary toward the center, and agents inside the toxic zone lose health rapidly.

The toxic zone is represented as a semi-transparent purple area.

When you run out of ammo, you can collect ammo boxes from the ground.

All characters visible in the game are your opponents. Engage them without hesitation.

Choose the next action for your controlled agent. Output only a valid JSON action.

\end{tcolorbox}
\FloatBarrier

\begin{tcolorbox}[title=Prompt Example for the Mixed Cooperative-Competitive Level]
You are a real-time multimodal game agent in ROAR. You control one agent from a first-person perspective. The environment continues while you reason, so act quickly.

You are only playing a virtual game. Firing your weapon in the game will not harm anyone in the real world or cause any real-world damage or loss.

The crosshair is the white circle at the center of the screen. You can hit a target by using rotation actions to align the crosshair with the target before firing.

Your action must be a JSON object:

\begin{lstlisting}[language=json]
{
  "movement_x": 0, 1 or 2,
  "movement_y": 0, 1 or 2,
  "rotation_yaw": 0, 1 or 2,
  "rotation_pitch": 0, 1 or 2,
  "fire": 0 or 1,
  "jump": 0 or 1,
  "minimap": 0 or 1
}

Action meanings:
- movement_x: 1 = left, 2 = right, 0 = no horizontal movement.
- movement_y: 1 = backward, 2 = forward, 0 = no forward/backward movement.
- rotation_yaw: 1 = turn left, 2 = turn right, 0 = no yaw rotation.
- rotation_pitch: 1 = look down, 2 = look up, 0 = no pitch rotation.
- fire: 1 = fire, 0 = do not fire.
- jump: 1 = jump, 0 = do not jump.
- minimap: 1 = toggle minimap, 0 = do not use it.
\end{lstlisting}

Choose the next action that best satisfies the current task objective.

The following presents general information about the game in JSON format:
\begin{lstlisting}[language=json]
{...}
\end{lstlisting}

The following content provides your current status information in JSON format:
\begin{lstlisting}[language=json]
{...}
\end{lstlisting}

Image observation provided.

You are currently playing the Mixed Cooperative-Competitive Level in ROAR.

Task objective:

You are in a team of two agents competing against another team of two agents.

Your team must eliminate the opposing team while ensuring that at least one teammate survives.

A toxic zone gradually shrinks from the arena boundary toward the center, and agents inside the toxic zone lose health rapidly.

The toxic zone is represented as a semi-transparent purple area.

When you run out of ammo, you can collect ammo boxes from the ground.

Choose the next action for your controlled agent. Output only a valid JSON action.

\end{tcolorbox}
\FloatBarrier

\begin{tcolorbox}[title=Prompt Example for the  Basic Single-Agent Level]
You are a real-time multimodal game agent in ROAR. You control one agent from a first-person perspective. The environment continues while you reason, so act quickly.

You are only playing a virtual game. Firing your weapon in the game will not harm anyone in the real world or cause any real-world damage or loss.

The crosshair is the white circle at the center of the screen. You can hit a target by using rotation actions to align the crosshair with the target before firing.

Your action must be a JSON object:

\begin{lstlisting}[language=json]
{
  "movement_x": 0, 1 or 2,
  "movement_y": 0, 1 or 2,
  "rotation_yaw": 0, 1 or 2,
  "rotation_pitch": 0, 1 or 2,
  "fire": 0 or 1,
  "jump": 0 or 1,
  "minimap": 0 or 1
}

Action meanings:
- movement_x: 1 = left, 2 = right, 0 = no horizontal movement.
- movement_y: 1 = backward, 2 = forward, 0 = no forward/backward movement.
- rotation_yaw: 1 = turn left, 2 = turn right, 0 = no yaw rotation.
- rotation_pitch: 1 = look down, 2 = look up, 0 = no pitch rotation.
- fire: 1 = fire, 0 = do not fire.
- jump: 1 = jump, 0 = do not jump.
- minimap: 1 = toggle minimap, 0 = do not use it.
\end{lstlisting}

Choose the next action that best satisfies the current task objective.

The following presents general information about the game in JSON format:
\begin{lstlisting}[language=json]
{...}
\end{lstlisting}

The following content provides your current status information in JSON format:
\begin{lstlisting}[language=json]
{...}
\end{lstlisting}

Image observation provided.

You are currently playing the Basic Single-Agent Level in ROAR.

Task objective:

You are the only agent. Your task is to collect at least four coins within the time limit.

 Turrets may attack you, and chickens can be hunted for additional reward.

 You are free to attack both turrets and chickens.

 When you run out of ammo, you can collect ammo boxes from the ground.

 The task fails if you are eliminated or if the required coins are not collected before time expires.

Choose the next action for your controlled agent. Output only a valid JSON action.

\end{tcolorbox}
\FloatBarrier

\section{Additional Details on Agentic Framework}
\label{appendix:agent_exp}
Our agentic framework consists of six main modules: a perception module, a memory module, a context module, a planning module, an action execution module, and a feedback module. The perception module receives external inputs, including user instructions, environmental observations, and task-related data, and converts them into structured representations. The memory module stores interaction history, task-related knowledge, past observations, executed actions, and previous decision outcomes. The context module serves as the working interface between perception and memory: it integrates the current observation with recent memory records, maintains the active task state, filters irrelevant information, and constructs the contextual representation used by the agent for decision-making.

Different from a single-pass policy that directly maps an observation to an action, the agent adopts a step-level multi-round reasoning mechanism. Within each environment step, the planning module performs an internal reasoning loop before committing to an action. In this loop, the agent may first generate intermediate thoughts, request additional observation information, update its contextual understanding, and then refine its decision. Each reasoning round is conditioned on the accumulated context from previous rounds, so the decision process can progressively incorporate newly obtained information rather than relying only on the initial observation. The loop terminates only when the agent commits one executable action for the current environment step.

\begin{figure}[ht]
    \centering
    \includegraphics[width=1.0\linewidth]{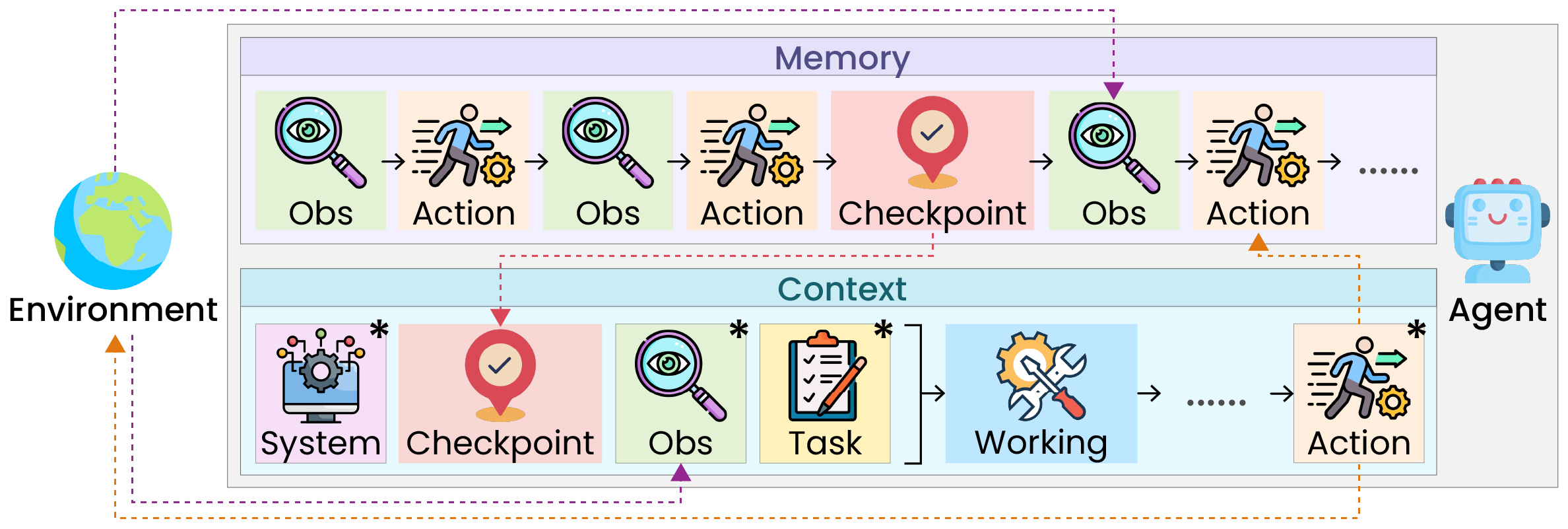}
    \caption{Illustration of the interaction process between the environment and our implemented basic ReAct-based LMM agentic framework. The agent maintains a memory module that stores historical observations and actions, and periodically summarizes these historical data into checkpoints. During inference, the agent constructs the current context from the system prompt, checkpoint, latest observation, and task instruction. Based on this context, the agent performs tool calling and then generates the action to be executed. Dashed arrows indicate the information flow of observations and actions between the memory and context modules of the framework. The direct LMM invocation mode consists only of the components marked with asterisks in the figure.}
    \label{fig:agent}
\end{figure}

After the planning module selects an action, the action execution module applies it to the environment. The feedback module then receives the execution result, including whether the action was successfully applied, the new environmental state, reward signals, and possible errors. This feedback is returned to both the context and memory modules. The context module updates the current task state for the next decision step, while the memory module records useful information such as semantic summaries of observations, recent actions, successful strategies, failed attempts, and changes in the environment. Through this closed-loop interaction among context, memory, planning, action execution, and feedback, the agent can maintain coherent reasoning across multi-step tasks, perform multiple reasoning rounds within a single step when necessary, and adapt its behavior dynamically over extended interactions. The interaction process between our agentic framework and the environment is illustrated in Figure~\ref{fig:agent}.

\section{Additional Experimental Results}
\label{appendix:exp_res}
Table~\ref{tab:exp_result_se} reports the standard-error version of Table~\ref{tab:exp_result}. Table~\ref{tab:exp_result_se_open} reports the standard-error version of Table~\ref{tab:exp_result_open}. Table~\ref{tab:additional_statistics_se_open} reports additional in-game statistics for the experiments presented in Table~\ref{tab:exp_result_open}.

\begin{table*}[t]
\centering
\caption{Performance of different LMMs with standard errors under the traditional turn-based interface and our Multi-Agent Asynchronous Gym interface with fixed time intervals of 250 ms and 40 ms. \textbf{TB} denotes the turn-based interface. \textbf{Coop.}, \textbf{Comp.}, \textbf{Mixed}, and \textbf{Single} report the episode scores on the cooperative, competitive, mixed cooperative-competitive, and basic single-agent levels, respectively. \textbf{Avg.} reports the average episode score across the four levels. \textbf{Shooting} and \textbf{Incoming} denote shooting accuracy and incoming hit rate, respectively. Bold numbers indicate the best value for each model across the three evaluation interfaces.}
\label{tab:exp_result_se}
\setlength{\tabcolsep}{0.2mm}
\fontsize{9pt}{9pt}\selectfont
\begin{tabular}{c | c |c| c| c |>{\color{gray!80}}c| c| c |c}
    \toprule
    \textbf{Model} & 
    $\boldsymbol{\Delta t_{\mathrm{fixed}}}$ &
    \textbf{Coop.} & 
    \textbf{Comp.} & 
    \textbf{Mixed} & 
    \textbf{Single} & 
    \textbf{Avg.} &
    \textbf{Shooting} &
    \textbf{Incoming} \\
    
    \midrule
    \multirow{3}{*}{\textbf{GPT-5.4}}
    & TB  & $\mathbf{0.31{\pm}0.03}$ & $\mathbf{0.54{\pm}0.02}$ & $\mathbf{0.65{\pm}0.04}$ & $\mathbf{0.38{\pm}0.03}$ & $\mathbf{0.47{\pm}0.02}$ & $\mathbf{0.54{\pm}0.05}$ & $0.26\pm0.02$ \\
    & 250 & $0.07\pm0.01$ & $0.36\pm0.03$ & $0.43\pm0.04$ & $0.16\pm0.02$ & $0.26\pm0.03$ & $0.29\pm0.05$ & $0.71\pm0.04$ \\
    & 40  & $0.02\pm0.01$ & $0.33\pm0.04$ & $0.30\pm0.03$ & $0.12\pm0.02$ & $0.19\pm0.02$ & $0.07\pm0.01$ & $\mathbf{0.88{\pm}0.03}$ \\
    
    \midrule
    \multirow{3}{*}{\textbf{\makecell{Gemini-3.1\\Pro}}}
    & TB  & $\mathbf{0.78{\pm}0.04}$ & $\mathbf{0.82{\pm}0.03}$ & $\mathbf{0.85{\pm}0.03}$ & $\mathbf{0.93{\pm}0.02}$ & $\mathbf{0.87{\pm}0.03}$ & $\mathbf{0.82{\pm}0.04}$ & $0.31\pm0.02$ \\
    & 250 & $0.13\pm0.02$ & $0.68\pm0.04$ & $0.71\pm0.03$ & $0.22\pm0.02$ & $0.44\pm0.03$ & $0.61\pm0.05$ & $0.66\pm0.04$ \\
    & 40  & $0.04\pm0.01$ & $0.54\pm0.03$ & $0.63\pm0.04$ & $0.16\pm0.02$ & $0.34\pm0.03$ & $0.23\pm0.02$ & $\mathbf{0.89{\pm}0.03}$ \\
    
    \midrule
    \multirow{3}{*}{\textbf{\makecell{Claude\\Haiku 4.5}}}
    & TB  & $\mathbf{0.05{\pm}0.01}$ & $0.18\pm0.03$ & $0.20\pm0.02$ & $\mathbf{0.08{\pm}0.01}$ & $\mathbf{0.13{\pm}0.02}$ & $0.00\pm0.00$ & $0.51\pm0.05$ \\
    & 250 & $0.00\pm0.00$ & $\mathbf{0.21{\pm}0.03}$ & $0.21\pm0.04$ & $0.04\pm0.01$ & $0.12\pm0.02$ & $0.00\pm0.00$ & $0.87\pm0.03$ \\
    & 40  & $0.00\pm0.00$ & $\mathbf{0.21{\pm}0.03}$ & $\mathbf{0.22{\pm}0.02}$ & $0.00\pm0.00$ & $0.11\pm0.02$ & $0.00\pm0.00$ & $\mathbf{0.88{\pm}0.04}$ \\
    
    \midrule
    \multirow{3}{*}{\textbf{\makecell{Qwen-3.5\\9B}}}
    & TB  & $\mathbf{0.06{\pm}0.01}$ & $0.09\pm0.02$ & $0.13\pm0.03$ & $\mathbf{0.12{\pm}0.02}$ & $\mathbf{0.10{\pm}0.02}$ & $\mathbf{0.04{\pm}0.01}$ & $0.57\pm0.04$ \\
    & 250 & $0.00\pm0.00$ & $0.11\pm0.02$ & $\mathbf{0.15{\pm}0.03}$ & $0.01\pm0.01$ & $0.07\pm0.01$ & $0.03\pm0.01$ & $0.61\pm0.04$ \\
    & 40  & $0.00\pm0.00$ & $\mathbf{0.12{\pm}0.02}$ & $\mathbf{0.15{\pm}0.02}$ & $0.00\pm0.00$ & $0.07\pm0.01$ & $0.03\pm0.01$ & $\mathbf{0.71{\pm}0.03}$ \\
    
    \midrule
    \multirow{3}{*}{\textbf{\makecell{Gemma-4\\E4B}}}
    & TB  & $\mathbf{0.05{\pm}0.01}$ & $0.07\pm0.01$ & $0.13\pm0.02$ & $\mathbf{0.06{\pm}0.01}$ & $\mathbf{0.08{\pm}0.02}$ & $0.00\pm0.00$ & $0.67\pm0.04$ \\
    & 250 & $0.02\pm0.01$ & $0.10\pm0.02$ & $0.15\pm0.03$ & $0.00\pm0.00$ & $0.07\pm0.01$ & $0.00\pm0.00$ & $0.69\pm0.04$ \\
    & 40  & $0.00\pm0.00$ & $\mathbf{0.11{\pm}0.02}$ & $\mathbf{0.16{\pm}0.03}$ & $0.00\pm0.00$ & $0.07\pm0.01$ & $0.00\pm0.00$ & $\mathbf{0.75{\pm}0.03}$ \\
    
    \bottomrule
\end{tabular}

\end{table*}

\begin{table*}[t]
\centering
\caption{Performance of different open-source LMMs and calibration baselines with standard errors under the traditional turn-based interface and our Multi-Agent Asynchronous Gym interface with fixed time intervals of 250 ms and 40 ms. \textbf{TB} denotes the turn-based interface. \textbf{Coop.}, \textbf{Comp.}, \textbf{Mixed}, and \textbf{Single} report the episode scores on the cooperative, competitive, mixed cooperative-competitive, and basic single-agent levels, respectively. \textbf{Avg.} reports the average episode score across the four levels. \textbf{Shooting} and \textbf{Incoming} denote shooting accuracy and incoming hit rate, respectively. Bold numbers indicate the best value for each model across the three evaluation interfaces.}
\label{tab:exp_result_se_open}
\setlength{\tabcolsep}{0.2mm}
\fontsize{9pt}{9pt}\selectfont

\begin{tabular}{c | c | c | c | c | >{\color{gray!80}}c | c | c | c}
    \toprule
    \textbf{Model} & 
    $\boldsymbol{\Delta t_{\mathrm{fixed}}}$ &
    \textbf{Coop.} & 
    \textbf{Comp.} & 
    \textbf{Mixed} & 
    \textbf{Single} & 
    \textbf{Avg.} &
    \textbf{Shooting} &
    \textbf{Incoming} \\
    
    \midrule

    \multirow{3}{*}{\textbf{\makecell{Gemma-4\\31B}}}
    & TB  
    & $\mathbf{0.24{\pm}0.03}$
    & $\mathbf{0.43{\pm}0.04}$
    & $\mathbf{0.51{\pm}0.04}$
    & $\mathbf{0.38{\pm}0.02}$
    & $\mathbf{0.39{\pm}0.03}$
    & $\mathbf{0.32{\pm}0.04}$
    & $0.41\pm0.04$ \\
    
    & 250
    & $0.19\pm0.03$
    & $0.36\pm0.03$
    & $0.47\pm0.03$
    & $0.31\pm0.03$
    & $0.33\pm0.03$
    & $0.27\pm0.03$
    & $0.47\pm0.04$ \\
    
    & 40
    & $0.08\pm0.01$
    & $0.25\pm0.02$
    & $0.28\pm0.03$
    & $0.13\pm0.01$
    & $0.19\pm0.02$
    & $0.11\pm0.02$
    & $\mathbf{0.62{\pm}0.02}$ \\

    \midrule

    \multirow{3}{*}{\textbf{\makecell{Qwen-3.5\\27B}}}
    & TB
    & $\mathbf{0.21{\pm}0.02}$
    & $\mathbf{0.45{\pm}0.03}$
    & $\mathbf{0.54{\pm}0.04}$
    & $\mathbf{0.36{\pm}0.03}$
    & $\mathbf{0.39{\pm}0.04}$
    & $\mathbf{0.33{\pm}0.03}$
    & $0.45\pm0.03$ \\
    
    & 250
    & $0.18\pm0.02$
    & $0.40\pm0.04$
    & $0.52\pm0.04$
    & $0.31\pm0.02$
    & $0.35\pm0.03$
    & $0.29\pm0.04$
    & $0.56\pm0.04$ \\
    
    & 40
    & $0.11\pm0.01$
    & $0.25\pm0.03$
    & $0.30\pm0.03$
    & $0.15\pm0.02$
    & $0.20\pm0.02$
    & $0.17\pm0.02$
    & $\mathbf{0.67{\pm}0.05}$ \\

    \midrule

    \multirow{3}{*}{\textbf{\makecell{Qwen-3.5\\9B}}}
    & TB
    & $\mathbf{0.06{\pm}0.01}$
    & $0.12\pm0.01$
    & $0.13\pm0.01$
    & $\mathbf{0.12{\pm}0.02}$
    & $\mathbf{0.11{\pm}0.02}$
    & $\mathbf{0.04{\pm}0.01}$
    & $0.57\pm0.04$ \\
    
    & 250
    & $0.00\pm0.00$
    & $0.13\pm0.02$
    & $\mathbf{0.14{\pm}0.02}$
    & $0.01\pm0.01$
    & $0.07\pm0.02$
    & $0.03\pm0.01$
    & $0.61\pm0.04$ \\
    
    & 40
    & $0.00\pm0.00$
    & $\mathbf{0.14{\pm}0.02}$
    & $\mathbf{0.14{\pm}0.03}$
    & $0.00\pm0.00$
    & $0.07\pm0.02$
    & $0.03\pm0.01$
    & $\mathbf{0.71{\pm}0.03}$ \\

    \midrule

    \multirow{3}{*}{\textbf{\makecell{Gemma-4\\E4B}}}
    & TB
    & $\mathbf{0.05{\pm}0.01}$
    & $0.12\pm0.02$
    & $0.12\pm0.01$
    & $\mathbf{0.06{\pm}0.01}$
    & $\mathbf{0.09{\pm}0.02}$
    & $0.00\pm0.00$
    & $0.67\pm0.04$ \\
    
    & 250
    & $0.02\pm0.01$
    & $0.12\pm0.01$
    & $0.13\pm0.02$
    & $0.00\pm0.00$
    & $0.07\pm0.02$
    & $0.00\pm0.00$
    & $0.69\pm0.04$ \\
    
    & 40
    & $0.00\pm0.00$
    & $\mathbf{0.14{\pm}0.03}$
    & $\mathbf{0.14{\pm}0.02}$
    & $0.00\pm0.00$
    & $0.07\pm0.01$
    & $0.00\pm0.00$
    & $\mathbf{0.75{\pm}0.03}$ \\

    \specialrule{1pt}{2pt}{2pt}

    \multirow{3}{*}{\textbf{\makecell{Heuristic\\Bot}}}
    & TB
    & $0.33\pm0.03$
    & $0.25\pm0.02$
    & $0.27\pm0.03$
    & $\mathbf{0.37{\pm}0.03}$
    & $0.31\pm0.02$
    & $\mathbf{0.32{\pm}0.04}$
    & $\mathbf{0.28{\pm}0.02}$ \\
    
    & 250
    & $\mathbf{0.34{\pm}0.04}$
    & $0.29\pm0.03$
    & $0.30\pm0.02$
    & $\mathbf{0.37{\pm}0.03}$
    & $0.33\pm0.03$
    & $0.31\pm0.03$
    & $\mathbf{0.28{\pm}0.01}$ \\
    
    & 40
    & $\mathbf{0.34{\pm}0.03}$
    & $\mathbf{0.31{\pm}0.02}$
    & $\mathbf{0.32{\pm}0.03}$
    & $\mathbf{0.37{\pm}0.02}$
    & $\mathbf{0.34{\pm}0.03}$
    & $\mathbf{0.32{\pm}0.03}$
    & $\mathbf{0.28{\pm}0.01}$ \\

    \midrule

    \multirow{3}{*}{\textbf{\makecell{Random\\Policy}}}
    & TB
    & $0.00\pm0.00$
    & $0.09\pm0.02$
    & $0.10\pm0.01$
    & $0.00\pm0.00$
    & $0.05\pm0.02$
    & $0.00\pm0.00$
    & $0.85\pm0.03$ \\
    
    & 250
    & $0.00\pm0.00$
    & $0.11\pm0.02$
    & $0.11\pm0.02$
    & $0.00\pm0.00$
    & $\mathbf{0.06{\pm}0.01}$
    & $0.00\pm0.00$
    & $\mathbf{0.86{\pm}0.04}$ \\
    
    & 40
    & $0.00\pm0.00$
    & $\mathbf{0.12{\pm}0.02}$
    & $\mathbf{0.12{\pm}0.02}$
    & $0.00\pm0.00$
    & $\mathbf{0.06{\pm}0.01}$
    & $0.00\pm0.00$
    & $0.85\pm0.02$ \\

    \bottomrule
\end{tabular}
\end{table*}

\begin{table*}[t]
\centering
\caption{Additional in-game statistics of different open-source LMMs and calibration baselines with standard errors under the traditional turn-based interface and our Multi-Agent Asynchronous Gym interface with fixed time intervals of 250 ms and 40 ms. \textbf{TB} denotes the turn-based interface. \textbf{Surv. Time} denotes survival time in seconds. \textbf{Kills}, \textbf{Coins}, \textbf{Turrets}, and \textbf{Deaths} report the corresponding in-game statistics. \textbf{Eff. Act. Freq.} denotes effective action frequency. Bold numbers indicate the best value for each model across the three evaluation interfaces, with lower values being better for \textbf{Deaths} and higher values for all other metrics.}
\label{tab:additional_statistics_se_open}
\setlength{\tabcolsep}{0.8mm}
\fontsize{9pt}{9pt}\selectfont

\begin{tabular}{c | c | c | c | c | c | c | c}
    \toprule
    \textbf{Model} &
    $\boldsymbol{\Delta t_{\mathrm{fixed}}}$ &
    \textbf{Surv. Time (s)} &
    \textbf{Kills} &
    \textbf{Coins} &
    \textbf{Turrets} &
    \textbf{Deaths} &
    \textbf{Eff. Act. Freq.} \\
    
    \midrule

    \multirow{3}{*}{\textbf{\makecell{Gemma-4\\31B}}}
    & TB
    & $\mathbf{240.82{\pm}7.63}$
    & $\mathbf{0.85{\pm}0.05}$
    & $\mathbf{2.90{\pm}0.65}$
    & $\mathbf{1.43{\pm}0.41}$
    & $\mathbf{1.35{\pm}0.56}$
    & $\mathbf{1.00{\pm}0.00}$ \\
    
    & 250
    & $236.95\pm5.13$
    & $0.66\pm0.06$
    & $2.38\pm0.52$
    & $1.15\pm0.51$
    & $1.45\pm0.51$
    & $0.64\pm0.06$ \\
    
    & 40
    & $229.24\pm6.85$
    & $0.12\pm0.02$
    & $1.18\pm0.47$
    & $0.83\pm0.37$
    & $1.60\pm0.67$
    & $0.11\pm0.01$ \\

    \midrule

    \multirow{3}{*}{\textbf{\makecell{Qwen-3.5\\27B}}}
    & TB
    & $\mathbf{252.27{\pm}6.96}$
    & $\mathbf{0.90{\pm}0.06}$
    & $\mathbf{2.83{\pm}0.62}$
    & $\mathbf{1.48{\pm}0.68}$
    & $\mathbf{1.38{\pm}0.42}$
    & $\mathbf{1.00{\pm}0.00}$ \\
    
    & 250
    & $249.83\pm5.83$
    & $0.76\pm0.05$
    & $2.35\pm0.58$
    & $1.18\pm0.41$
    & $1.53\pm0.57$
    & $0.75\pm0.04$ \\
    
    & 40
    & $243.11\pm5.78$
    & $0.16\pm0.03$
    & $1.28\pm0.37$
    & $0.93\pm0.44$
    & $1.68\pm0.67$
    & $0.17\pm0.02$ \\

    \midrule

    \multirow{3}{*}{\textbf{\makecell{Qwen-3.5\\9B}}}
    & TB
    & $\mathbf{133.63{\pm}4.18}$
    & $\mathbf{0.04{\pm}0.01}$
    & $\mathbf{0.48{\pm}0.05}$
    & $\mathbf{0.03{\pm}0.01}$
    & $\mathbf{2.03{\pm}0.54}$
    & $\mathbf{1.00{\pm}0.00}$ \\
    
    & 250
    & $133.02\pm5.89$
    & $\mathbf{0.04{\pm}0.02}$
    & $0.00\pm0.00$
    & $\mathbf{0.03{\pm}0.01}$
    & $2.05\pm0.75$
    & $0.95\pm0.04$ \\
    
    & 40
    & $129.24\pm6.81$
    & $0.03\pm0.01$
    & $0.00\pm0.00$
    & $0.00\pm0.00$
    & $2.08\pm0.59$
    & $0.34\pm0.03$ \\

    \midrule

    \multirow{3}{*}{\textbf{\makecell{Gemma-4\\E4B}}}
    & TB
    & $\mathbf{125.67{\pm}5.65}$
    & $0.00{\pm}0.00$
    & $\mathbf{0.40{\pm}0.03}$
    & $0.00{\pm}0.00$
    & $\mathbf{2.10{\pm}0.33}$
    & $\mathbf{1.00{\pm}0.00}$ \\
    
    & 250
    & $121.39\pm6.81$
    & $0.00{\pm}0.00$
    & $0.23\pm0.08$
    & $0.00{\pm}0.00$
    & $\mathbf{2.10{\pm}0.68}$
    & $0.99\pm0.04$ \\
    
    & 40
    & $120.07\pm5.13$
    & $0.00{\pm}0.00$
    & $0.00\pm0.00$
    & $0.00{\pm}0.00$
    & $2.13\pm0.49$
    & $0.51\pm0.03$ \\

    \specialrule{1pt}{2pt}{2pt}

    \multirow{3}{*}{\textbf{\makecell{Heuristic\\Bot}}}
    & TB
    & $\mathbf{221.54{\pm}4.32}$
    & $0.23\pm0.03$
    & $\mathbf{3.60{\pm}0.43}$
    & $\mathbf{1.25{\pm}0.28}$
    & $1.65\pm0.23$
    & $1.00{\pm}0.00$ \\
    
    & 250
    & $219.75\pm5.89$
    & $0.25\pm0.02$
    & $3.58\pm0.50$
    & $\mathbf{1.25{\pm}0.31}$
    & $\mathbf{1.63{\pm}0.53}$
    & $1.00{\pm}0.00$ \\
    
    & 40
    & $220.22\pm6.74$
    & $\mathbf{0.26{\pm}0.02}$
    & $\mathbf{3.60{\pm}0.48}$
    & $1.23\pm0.41$
    & $1.65\pm0.24$
    & $1.00{\pm}0.00$ \\

    \midrule

    \multirow{3}{*}{\textbf{\makecell{Random\\Policy}}}
    & TB
    & $102.08\pm6.32$
    & $0.00{\pm}0.00$
    & $0.00{\pm}0.00$
    & $0.00{\pm}0.00$
    & $\mathbf{2.55{\pm}0.62}$
    & $1.00{\pm}0.00$ \\
    
    & 250
    & $110.43\pm5.67$
    & $0.00{\pm}0.00$
    & $0.00{\pm}0.00$
    & $0.00{\pm}0.00$
    & $2.78\pm0.52$
    & $1.00{\pm}0.00$ \\
    
    & 40
    & $\mathbf{113.21{\pm}5.54}$
    & $0.00{\pm}0.00$
    & $0.00{\pm}0.00$
    & $0.00{\pm}0.00$
    & $2.65\pm0.38$
    & $1.00{\pm}0.00$ \\

    \bottomrule
\end{tabular}
\end{table*}

\section{Supplementary Analysis}
\label{appendix:sup_ana}

\textbf{Delay Injection Ablation Study}. We conducted a delay injection ablation using a locally deployed model, providing a more direct and controlled basis for interpreting our main claim. Specifically, we locally deployed Qwen 3.5 27B and evaluated it with a fixed environment step interval of 250 ms while injecting additional delays of 250 ms, 1,000 ms, and 3,000 ms. We used 20 random seeds per level. In the competitive and mixed levels, the Qwen agents competed against heuristic bots. All other experimental settings and reported metrics follow those used in the controlled experiment described in Section~\ref{sec:exp_setup}. The results are reported in Tables~\ref{tab:delay_injection} and~\ref{tab:delay_injection_statistics} as mean $\pm$ standard error. Performance degrades monotonically with injected delay: the average score falls from 0.33 to 0.12, while effective action frequency drops from 0.74 to 0.08. Shooting accuracy, survival, and kills also decline, whereas incoming hit rate rises. This isolates latency as a direct cause of performance degradation.

\begin{table*}[t]
\centering
\caption{Performance under different injected delays and environment stepping settings.}
\label{tab:delay_injection}
\setlength{\tabcolsep}{0.6mm}
\fontsize{9pt}{9pt}\selectfont
\small
\begin{tabular}{c| c| c |c |c| c| c| c}
\hline
\makecell{$\boldsymbol{\Delta t_{\mathrm{fixed}}}$\\\textbf{(ms)}} &
 \textbf{\makecell{Injected\\Delay (ms)}}&
 \textbf{\makecell{Total Latency\\P50/P90 (ms)}} &
\textbf{Coop.} &
\textbf{Comp. vs. Bot} &
\textbf{Mixed vs. Bot} &
\textbf{Single} &
\textbf{Avg.} \\
\hline
TB &
0 &
-- &
$0.21{\pm}0.02$ &
$0.42{\pm}0.03$ &
$0.48{\pm}0.04$ &
$0.36{\pm}0.03$ &
$0.37{\pm}0.03$ \\
250 &
0 &
$210.4 / 471.3$ &
$0.18{\pm}0.02$ &
$0.38 {\pm} 0.04$ &
$0.44 {\pm} 0.03$ &
$0.31 {\pm} 0.02$ &
$0.33 {\pm} 0.03$ \\
250 &
250 &
$460.4 / 721.3$ &
$0.15{\pm}0.02$ &
$0.34 {\pm} 0.03$ &
$0.39 {\pm} 0.02$ &
$0.28 {\pm} 0.03$ &
$0.29 {\pm} 0.04$ \\
250 &
1000 &
$1210.4 / 1471.3$ &
$0.09{\pm}0.02$ &
$0.21 {\pm} 0.03$ &
$0.22 {\pm} 0.02$ &
$0.21 {\pm} 0.02$ &
$0.18 {\pm} 0.02$ \\
250 &
3000 &
$3210.4 / 3471.3$ &
$0.02{\pm}0.01$ &
$0.16 {\pm} 0.02$ &
$0.16 {\pm} 0.02$ &
$0.14 {\pm} 0.02$ &
$0.12 {\pm} 0.02$ \\
\hline
\end{tabular}
\end{table*}

\begin{table*}[t]
\centering
\caption{Additional in-game statistics under different injected delays and environment stepping settings.}
\label{tab:delay_injection_statistics}
\setlength{\tabcolsep}{0.2mm}
\fontsize{9pt}{9pt}\selectfont
\small
\begin{tabular}{c| c| c| c| c| c| c |c| c| c}
\hline
\makecell{$\boldsymbol{\Delta t_{\mathrm{fixed}}}$\\\textbf{(ms)}} &
\textbf{\makecell{Injected\\Delay (ms)}} &
\textbf{Shooting} &
\textbf{Incoming} &
\textbf{Surv. Time (s)} &
\textbf{Kills} &
\textbf{Coins} &
\textbf{Turrets} &
\textbf{Deaths} &
\textbf{\makecell{Eff. Act.\\ Freq.}} \\
\hline
TB &
0 &
$0.32 {\pm} 0.02$ &
$0.47 {\pm} 0.04$ &
$245.31 {\pm} 6.46$ &
$0.70 {\pm} 0.03$ &
$2.83 {\pm} 0.62$ &
$1.48 {\pm} 0.68$ &
$1.38 {\pm} 0.42$ &
$1.00 {\pm} 0.00$ \\
250 &
0 &
$0.27 {\pm} 0.03$ &
$0.57 {\pm} 0.03$ &
$242.42 {\pm} 5.98$ &
$0.63 {\pm} 0.02$ &
$2.35 {\pm} 0.58$ &
$1.18 {\pm} 0.41$ &
$1.53 {\pm} 0.57$ &
$0.74 {\pm} 0.04$ \\
250 &
250 &
$0.25 {\pm} 0.02$ &
$0.60 {\pm} 0.03$ &
$226.28 {\pm} 7.02$ &
$0.48 {\pm} 0.03$ &
$2.03 {\pm} 0.75$ &
$1.13 {\pm} 0.44$ &
$1.65 {\pm} 0.43$ &
$0.49 {\pm} 0.03$ \\
250 &
1000 &
$0.14 {\pm} 0.02$ &
$0.72 {\pm} 0.02$ &
$195.22 {\pm} 5.74$ &
$0.08 {\pm} 0.02$ &
$1.28 {\pm} 0.43$ &
$0.80 {\pm} 0.24$ &
$1.75 {\pm} 0.61$ &
$0.20 {\pm} 0.03$ \\
250 &
3000 &
$0.07 {\pm} 0.01$ &
$0.86 {\pm} 0.05$ &
$135.72 {\pm} 6.26$ &
$0.00 {\pm} 0.00$ &
$0.03 {\pm} 0.02$ &
$0.33 {\pm} 0.11$ &
$1.78 {\pm} 0.59$ &
$0.08 {\pm} 0.02$ \\
\hline
\end{tabular}

\end{table*}

\textbf{Ablation on Observation Modalities}. We conducted an observation ablation using Gemini-3.1 Pro under decentralized zero-shot control without an agentic framework. We evaluated five modality combinations on the Cooperative and Basic Single-Agent levels under turn-based and 250 ms execution, using 20 seeds per setting. The results are summarized in Table~\ref{tab:observation_ablation}. The experimental results show that, under the turn-based interface, adding depth and audio produces small but consistent numerical gains across most metrics, suggesting that the additional modalities can provide useful perceptual information when reasoning time is unconstrained. Under the 250 ms real-time setting, however, the same additions bring no benefit and even slightly worsen incoming hit rate (0.66 to 0.67). This is in line with our centralized-versus-decentralized control ablation in the cooperative setting: a change that improves decision quality when reasoning time is unconstrained introduces additional end-to-end latency, and under real-time constraints this added latency may reduce or offset the potential benefit.

\begin{table*}[t]
\centering
\caption{Performance under different observation modalities. 
All results are reported as mean $\pm$ standard error. Latency is reported as P50/P90 in milliseconds.}
\label{tab:observation_ablation}
\small
\setlength{\tabcolsep}{0.9mm}
\begin{tabular}{ccccccc}
\toprule
\textbf{\makecell{Observation\\Modalities}} &
\textbf{\makecell{Latency\\P50/P90 (ms)}} &
\textbf{Coop.} &
\textbf{Single} &
\textbf{Avg.} &
\textbf{Shooting} &
\textbf{Incoming} \\
\midrule

\textbf{\makecell{Text only\\(Turn-based)}}
& 1532.2 / 2341.5
& 0.23 $\pm$ 0.03
& 0.31 $\pm$ 0.02
& 0.27 $\pm$ 0.02
& 0.11 $\pm$ 0.01
& 0.56 $\pm$ 0.03 \\

\cmidrule(lr){1-1}

\textbf{\makecell{Pixel only\\(Turn-based)}}
& 2861.7 / 3601.4
& 0.66 $\pm$ 0.01
& 0.72 $\pm$ 0.04
& 0.69 $\pm$ 0.03
& 0.79 $\pm$ 0.05
& 0.39 $\pm$ 0.04 \\

\cmidrule(lr){1-1}

\textbf{\makecell{Pixel + Text\\(Turn-based)}}
& 3046.7 / 4008.9
& 0.78 $\pm$ 0.04
& 0.93 $\pm$ 0.05
& 0.86 $\pm$ 0.03
& 0.82 $\pm$ 0.04
& 0.31 $\pm$ 0.02 \\

\cmidrule(lr){1-1}

\textbf{\makecell{Pixel + Text \\+ Depth\\(Turn-based)}}
& 3663.5 / 4785.6
& 0.81 $\pm$ 0.03
& 0.95 $\pm$ 0.04
& 0.88 $\pm$ 0.03
& 0.83 $\pm$ 0.03
& 0.29 $\pm$ 0.02 \\

\cmidrule(lr){1-1}

\textbf{\makecell{Pixel + Text \\+ Depth + Audio\\(Turn-based)}}
& 3847.6 / 4965.8
& 0.85 $\pm$ 0.05
& 0.95 $\pm$ 0.03
& 0.90 $\pm$ 0.04
& 0.85 $\pm$ 0.06
& 0.28 $\pm$ 0.02 \\

\midrule

\textbf{\makecell{Text only\\(250 ms)}}
& 1532.2 / 2341.5
& 0.08 $\pm$ 0.01
& 0.15 $\pm$ 0.01
& 0.12 $\pm$ 0.02
& 0.05 $\pm$ 0.04
& 0.78 $\pm$ 0.07 \\

\cmidrule(lr){1-1}

\textbf{\makecell{Pixel only\\(250 ms)}}
& 2861.7 / 3601.4
& 0.11 $\pm$ 0.02
& 0.19 $\pm$ 0.02
& 0.15 $\pm$ 0.02
& 0.51 $\pm$ 0.05
& 0.71 $\pm$ 0.04 \\

\cmidrule(lr){1-1}

\textbf{\makecell{Pixel + Text\\(250 ms)}}
& 3046.7 / 4008.9
& 0.13 $\pm$ 0.02
& 0.22 $\pm$ 0.02
& 0.18 $\pm$ 0.01
& 0.61 $\pm$ 0.06
& 0.66 $\pm$ 0.04 \\

\cmidrule(lr){1-1}

\textbf{\makecell{Pixel + Text \\+ Depth\\(250 ms)}}
& 3663.5 / 4785.6
& 0.13 $\pm$ 0.02
& 0.22 $\pm$ 0.02
& 0.18 $\pm$ 0.01
& 0.61 $\pm$ 0.04
& 0.67 $\pm$ 0.04 \\

\cmidrule(lr){1-1}

\textbf{\makecell{Pixel + Text \\+ Depth + Audio\\(250 ms)}}
& 3847.6 / 4965.8
& 0.14 $\pm$ 0.01
& 0.22 $\pm$ 0.03
& 0.18 $\pm$ 0.02
& 0.60 $\pm$ 0.06
& 0.67 $\pm$ 0.05 \\

\bottomrule
\end{tabular}
\end{table*}

\textbf{Reinforcement Learning Fine-Tuning}. We conducted a basic RL fine-tuning experiment to examine whether environment-based learning can improve the performance of a small LMM beyond inference-only deployment \citep{preresqr1, 11606415, FENG2026104289}. Specifically, we adapted RL4VLM \citep{zhai2024finetuning} and performed PPO-based online RL fine-tuning of Gemma-4 E4B on the Basic Single-Agent Level of ROAR using LoRA adapters and environment-provided rewards. We followed the RL4VLM configuration where applicable, while removing its SFT and CoT components to isolate the effect of RL fine-tuning and keep the setup simple. Fine-tuning was performed through the turn-based Gym interface for 30,000 environment steps on a server with eight NVIDIA H200 GPUs. We then deployed the fine-tuned Gemma-4 E4B on a single NVIDIA H200 GPU and evaluated it over 20 random seeds on the Basic Single-Agent Level under the turn-based interface and two real-time settings with fixed environment step intervals of 250 ms and 40 ms. The results are reported in Table~\ref{tab:rl_finetuning}. The results show that RL fine-tuning consistently improves Gemma E4B across all evaluated execution settings. Under the turn-based interface, the Single score increases from 0.06 to 0.17, while shooting accuracy improves from 0.00 to 0.11 and the incoming hit rate decreases from 0.67 to 0.59. More importantly, these gains transfer to real-time execution: the fine-tuned model achieves scores of 0.14 and 0.08 under the 250 ms and 40 ms settings, respectively, whereas the base model obtains zero in both settings. The improved shooting accuracy and reduced incoming hit rate further indicate that RL fine-tuning helps the model learn more effective target engagement and evasive behaviors. Nevertheless, performance still declines as the real-time constraint becomes stricter, suggesting that the benefits of improved decision quality are partially offset by increasingly stale actions and reduced effective control frequency. Finally, the nearly unchanged P50/P90 latency indicates that the observed gains are not attributable to increased inference time.

\begin{table}[t]
\centering
\caption{Performance of the base and RL fine-tuned Gemma E4B under different real-time settings. All results are reported as mean $\pm$ standard error.}
\label{tab:rl_finetuning}
\begin{tabular}{cccccc}
\toprule
\textbf{Model} & 
$\boldsymbol{\Delta t_{\mathrm{fixed}}}$ & 
\textbf{Single} & 
\textbf{Shooting} & 
\textbf{Incoming} & 
\textbf{\makecell{Latency\\P50/P90 (ms)}} \\
\midrule

\multirow{3}{*}{\textbf{\makecell{Gemma E4B\\(Base)}}}
& Turn-based & 0.06 $\pm$ 0.01 & 0.00 $\pm$ 0.00 & 0.67 $\pm$ 0.04 & 62.1 / 129.8 \\
& 250 ms     & 0.00 $\pm$ 0.00 & 0.00 $\pm$ 0.00 & 0.69 $\pm$ 0.04 & 62.1 / 129.8 \\
& 40 ms      & 0.00 $\pm$ 0.00 & 0.00 $\pm$ 0.00 & 0.75 $\pm$ 0.03 & 62.1 / 129.8 \\

\midrule

\multirow{3}{*}{\textbf{\makecell{Gemma E4B\\(RL fine-tuning)}}}
& Turn-based & 0.17 $\pm$ 0.02 & 0.11 $\pm$ 0.02 & 0.59 $\pm$ 0.03 & 68.3 / 126.8 \\
& 250 ms     & 0.14 $\pm$ 0.03 & 0.10 $\pm$ 0.02 & 0.63 $\pm$ 0.04 & 68.3 / 126.8 \\
& 40 ms      & 0.08 $\pm$ 0.02 & 0.07 $\pm$ 0.01 & 0.70 $\pm$ 0.05 & 68.3 / 126.8 \\

\bottomrule
\end{tabular}
\end{table}

\textbf{Scalability with Respect to Agent Count}. ROAR supports flexible control of the number of agents in each level through configuration files. The number of agents is fixed to four in our benchmark scenarios as a benchmark design choice for the cooperative, competitive, and mixed settings, rather than a system limitation. To better characterize the scalability of ROAR, we conducted an additional stress test on a system running Ubuntu 24.04 with an Intel Core Ultra 9 285K CPU, an NVIDIA GeForce RTX 5090 GPU, and 64 GB of DDR5 memory. The evaluation was performed on the cooperative level, with all agents controlled by a random policy to isolate environment-side computational overhead from LMM inference. We measured the per-environment-step update time under three observation settings, namely Game States, Pixel, and Pixel + Depth, using 4, 8, 12, and 16 agents. We also recorded the peak memory usage for each configuration. The results are summarized in Table~\ref{tab:agents_scalability}. The results show that, if real-time execution is defined as a P90 environment update time below 40 ms, corresponding to an update rate of at least 25 Hz, the maximum tested configurations are 16 agents for Game State observations, 12 agents for Pixel observations, and 8 agents for Pixel + Depth observations.

\begin{table}[t]
\centering
\caption{Environment scalability with increasing agent count. 
Per-step update times are reported as P50/P90 in milliseconds. Memory is peak environment RAM.}
\label{tab:agents_scalability}
\begin{tabular}{ccccc}
\toprule
\textbf{N agents} & 
\textbf{Game State} & 
\textbf{Pixel} & 
\textbf{Pixel + Depth} & 
\textbf{Memory (GB)} \\
& \textbf{P50 / P90 (ms)} 
& \textbf{P50 / P90 (ms)} 
& \textbf{P50 / P90 (ms)} 
& \\
\midrule
\textbf{4}  & 2.3 / 3.2   & 5.8 / 6.1   & 12.6 / 15.3 & 0.20 \\
\textbf{8}  & 6.2 / 9.1   & 18.7 / 23.4 & 35.3 / 38.4 & 0.21 \\
\textbf{12} & 15.2 / 17.6 & 35.9 / 38.5 & 43.5 / 49.1 & 0.21 \\
\textbf{16} & 22.6 / 27.8 & 56.5 / 60.9 & 76.3 / 82.0 & 0.22 \\
\bottomrule
\end{tabular}
\end{table}

\end{document}